\documentclass[aps,prd,preprint,groupedaddress]{revtex4-1}
\usepackage{extraipa}
\usepackage{amsmath}
\usepackage{amsfonts}
\usepackage{amssymb}

\newcommand{\etal}{\textit{et al}.}

\begin{document}

\title{Quantum aspects of antisymmetric tensor field with spontaneous Lorentz violation}

\author{Sandeep Aashish}%
\email[]{sandeepa16@iiserb.ac.in}

\author{Sukanta Panda}
\email[]{sukanta@iiserb.ac.in}

\affiliation{Department of Physics, Indian Institute of Science Education and Research, Bhopal 462066, India}

%


\date{\today}

\begin{abstract}
We study the quantization of a simple model of antisymmetric tensor field with spontaneous Lorentz violation in curved spacetime. We evaluate the 1-loop corrections at first order of metric perturbation, using a general covariant effective action approach. We revisit the issue of quantum equivalence, and find that it holds for non-Lorentz-violating modes but breaks down for Lorentz violating modes.
\end{abstract}



\maketitle

\section{\label{intro}Introduction}
The quest for quantizing gravity is ultimately related to understanding physics at the planck scale, candidates for which include string theory and loop quantum gravity. A difficulty that the development of such theories faces, is our inability to probe high energy scales, owing to the limitations of current particle physics experiments \cite{zimmermann2018}. This has led to significant efforts towards finding low energy signatures using effective field theory tools that could be relevant in current and near future experiments in both particle physics and early universe cosmology. Phenomenologically, this amounts to detecting planck supressed variations to standard model and general relativity while maintaining observer independence, termed as standard model extension (SME) \cite{kostelecky1995,colladay1997,*colladay1998,kostelecky2004,bluhm2006}. 

There is substantial evidence of SME effects from string theory and quantum gravity, according to which certain mechanisms could lead to violation of Lorentz symmetry \cite{kostelecky1989b,*kostelecky1991a,kostelecky1991b,*kostelecky1996,*kostelecky2001,gambini1999,alfaro2002, sudarsky2002,*sudarsky2003,myers2003}, which is a fundamental symmetry in general relativity that relates all physical local Lorentz frames. In principle, Lorentz violation can be introduced in a theory either explicitly, in which case the Lagrange density is not Lorentz invariant, or spontaneously, so that the Lagrange density is Lorentz invariant but the physics can still display Lorentz violation \cite{kostelecky1991a,kostelecky1989b}. However, theories with explicit Lorentz violation have been found to be problematic due to their incompatibility with Bianchi identities in Riemann geometry \cite{kostelecky2004}, and are therefore not favourable for studies involving gravity.

Another consequence of string theory, at low energies, is the appearance of antisymmetric tensor field along with a symmetric tensor (metric) and a dilaton (scalar field) as a result of compactification of higher dimensions \cite{rohm1986,ghezelbash2009}. Until recently, antisymmetric tensor had not recieved serious consideration in studies of early universe cosmology, in particular inflation, due to some generic instability issues \cite{koivisto2009a,aashish2018c,prokopec2006}, but some recent studies have shown that presence of antisymmetric tensor field is likely to play a role during inflation era \cite{aashish2019a,elizalde2018}. Hence, as a natural extension, an interesting exercise is to consider Lorentz violation in conjunction with antisymmetric tensor (see, for example Ref. \cite{petrov2019}).

Altschul \etal \ in Ref. \cite{altschul2010} explored in detail spontaneous Lorentz violation with antisymmetric tensor fields, and found the presence of distinctive physical features with phenomenological implications for tests of Lorentz violation, even with relatively simple antisymmetric field models with a gauge invariant kinetic term. 

Our interest in the present work is to take first steps to extend the classical analysis in Ref. \cite{altschul2010} to quantum regime. We focus on the formal aspects of quantization of antisymmetric tensor field with spontaneous Lorentz violation, and primarily restrict ourselves to dealing with two issues. First, we set up the framework to evaluate the one-loop effective action using covariant effective action approach \cite{dewitt1964,dewitt1967a,dewitt1967b,dewitt1967c,vilkovisky1984a,vilkovisky1984b,parker2009}. For simplicity, we consider an action with only quadratic order terms, but in a \textit{nearly flat} spacetime (Minkowski metric $\eta_{\mu\nu}$ plus a classical perturbation $\kappa h_{\mu\nu}$). This yields one-loop corrections at $O(\kappa \hbar)$, involving terms upto first order in $h_{\mu\nu}$. Second, we check the quantum equivalence of the quadratic action considered in the first part with a classically equivalent vector theory, at 1-loop level. The issue of quantum equivalence in curved spacetime is interesting because a free massive antisymmetric tensor theory (no Lorentz violation) is known to be equivalent to a massive vector theory at classical and quantum level due to topological properties of zeta functions \cite{buchbinder2008} but, such properties do not hold when Lorentz symmetry is spontaneously broken \cite{aashish2018b}. In fact, it was demonstrated by Seifert in Ref. \cite{seifert2010a} that interaction of vector and tensor theories with gravity are different when topologically nontrivial monopole-like solutions of the spontaneous symmetry breaking equations exist. The method presented here is quite general in terms of its applicability to models with higher order terms in fields. 

In Sec. \ref{sec2}, we briefly review spontaneous Lorentz violation in antisymmetric tensor and introduce the classical action considered in this work. The notations used here are largely inspired by Ref. \cite{altschul2010}. We discuss the covariant effective action technique and its application to derive 1-loop corrections in Sec. \ref{sec3}. We also calculate the various propagators required to solve the 1-loop integrals. In Sec. \ref{sec4}, we consider the classically equivalent vector theory and calculate 1-loop corrections to compare with the results of Sec. \ref{sec3}, to check the quantum equivalence. 

\section{\label{sec2}Spontaneous Lorentz Violation and classical action}
Spontaneous symmetry breaking occurs when the equations of motion obey a symmetry but the solutions do not, and is effected via fixing a preferred value of vacuum (ground state) solutions. In general relativity, physically equivalent coordinate (or observer) frames are related via general coordinate transformations and local Lorentz transformations. In a given observer frame, fixing the vacuum expectation value (vev) of a tensor or vector field leads to spontaneous breaking of Lorentz symmetry, since all couplings with vev have preferred directions in spacetime \cite{bluhm2005,bluhm2006}. 

Spontaneous Lorentz violation in tensor field Lagrangians can be introduced through a potential term that drives a nonzero vacuum value of tensor field. For an antisymmetric 2-tensor field $B_{\mu\nu}$, we assume,
\begin{eqnarray}
\label{slva0}
\langle B_{\mu\nu}\rangle = b_{\mu\nu}.
\end{eqnarray}
It is possible to attain a special observer frame in a local Lorentz frame in Riemann spacetime or everywhere in Minkowski spacetime, in which $b_{\mu\nu}$ takes a simple block-diagonal form \cite{altschul2010},
\begin{eqnarray}
\label{slva1}
b_{\mu\nu} = 
\left(\begin{matrix}
0 & -a & 0 & 0\\
a & 0 & 0 & 0\\
0 & 0 & 0 & b\\
0 & 0 & -b & 0
\end{matrix}\right),
\end{eqnarray}
provided at least one of the quantities $X_{1}=-2(a^2-b^2)$ and $X_{2}=4ab$ is nonzero, where $a$ and $b$ are real numbers. Moreover, the analysis of monopole solutions of antisymmetric tensor in Ref. \cite{seifert2010b} showed that for a spherically symmetric nontrivial solution of the equation of motion of $B_{\mu\nu}$ that asymptotically approaches vev, the potential of the form considered below (Eq. (\ref{slva2})) requires putting $a=0$. As will be seen later on, this choice of $a$ also ensures positivity of certain determinants appearing in loop integral calculations. We thus assume $a=0$ in the present analysis, although most of the calculations presented here are independent of the structure of $b_{\mu\nu}$. For later convenience, we also choose $b_{\mu\nu}b^{\mu\nu}=1$. 

We consider a simple model of a rank-2 antisymmetric tensor field, $B_{\mu\nu}$, with a spontaneous Lorentz violation inducing potential \cite{altschul2010},  
\begin{eqnarray}
\label{slva2}
V(B) = \frac{1}{16}\alpha^2 \Big(B_{\mu\nu}B^{\mu\nu} - b_{\mu\nu}b^{\mu\nu}\Big)^{2}.
\end{eqnarray}
Again, for the purpose of present analysis, we would like to consider only quadratic order terms in $B_{\mu\nu}$. To this end, we consider small fluctuations of $B_{\mu\nu}$ about a background value $b_{\mu\nu}$ \cite{altschul2010}, 
\begin{eqnarray}
\label{slva3}
B_{\mu\nu} = b_{\mu\nu} + \tilde{B}_{\mu\nu}.
\end{eqnarray} 
and neglect quartic and cubic terms in fluctuations $\tilde{B}_{\mu\nu}$ assuming $|| \tilde{B}_{\mu\nu}|| \ll ||b_{\mu\nu}||$. The resulting potential is,
\begin{eqnarray}
\label{slva4}
V(B) \approx \dfrac{1}{4}\alpha^2\Big(b_{\mu\nu}\tilde{B}^{\mu\nu}\Big)^2.
\end{eqnarray}
Although it may seem at this point that a quadratic Lagrangian might not lead to any significant physical result upon quantization, and it is actually true in case of a flat spacetime, nontrivial physical contributions appear in the 1-loop effective action in curved spacetime as demonstrated in the next section. For notational convenience, we do not explicitly write the \textit{tilde} symbol for field fluctuations, and assume its use throughout. We thus work with the Lagrangian,
\begin{eqnarray}
\label{slva5}
\mathcal{L} = -\frac{1}{12}H_{\mu\nu\lambda}H^{\mu\nu\lambda} - \dfrac{1}{4}\alpha^{2} \Big(b_{\mu\nu}{B}^{\mu\nu}\Big)^{2}.
\end{eqnarray}
The first term in Eq. (\ref{slva5}) is the gauge invariant kinetic term, 
\begin{equation}
\label{slva6}
H_{\mu\nu\lambda} \equiv \nabla_{\mu}B_{\nu\lambda} + \nabla_{\lambda}B_{\mu\nu} + \nabla_{\nu}B_{\lambda\mu},
\end{equation}
obeying the symmetry: $B^{\mu\nu} \longrightarrow B^{\mu\nu} + \nabla_{\mu}\xi_{\nu} - \nabla_{\nu}\xi_{\mu}$ for a gauge parameter $\xi_{\mu}$. The gauge invariance of kinetic term in an otherwise non-gauge invariant Lagrangian (\ref{slva5}) gives rise to redundancy problems in the energy spectrum \cite{buchbinder2007}, and cannot be removed via usual quantization method. A consistent method to treat this redundancy is given by the St{\"u}ckelberg procedure \cite{stuckelberg1957}. According to this procedure, a strongly coupled field called the St{\"u}ckelberg field is introduced in the symmetry breaking potential term such that the gauge symmetry is restored in a given Lagrangian. The original theory is still recovered in a special gauge (where St{\"u}ckelberg field is put to zero), however, the advantage is that the redundant degrees of freedom are now encompassed in the St{\"u}ckelberg field, and can be dealt with using well known quantization frameworks like the Faddeev-Popov method. For a detailed account of this procedure applied to massless and massive antisymmetric tensors, interested reader is referred to Refs. \cite{buchbinder1992,buchbinder2008} respectively, and to Ref. \cite{aashish2018a} for a more recent analysis in the context of covariant effective action.  

The above procedure is applied to (\ref{slva5}) via the introduction of a strongly coupled vector field $C_{\mu}$:
\begin{eqnarray}
\label{slva7}
\mathcal{L} = -\frac{1}{12}H_{\mu\nu\lambda}H^{\mu\nu\lambda} - \frac{1}{4}\alpha^{2}\Big[b_{\mu\nu}\Big(B^{\mu\nu} + \frac{1}{\alpha}F^{\mu\nu}[C]\Big)\Big]^{2},
\end{eqnarray}
so that the Lagrangian (\ref{slva7}) becomes gauge invariant (here, $F_{\mu\nu}\equiv\partial_{\mu}C_{\nu}-\partial_{\nu}C_{\mu}$), and reduces to original Lagrangian (\ref{slva5}) in the gauge $C_{\mu}=0$. The new Lagrangian is invariant under two sets of transformations: $(i)$ gauge transformation of $B_{\mu\nu}$ and shift of field $C_{\mu}$, 
\begin{eqnarray}
\label{slvb0}
B^{\mu\nu} &\longrightarrow & B^{\mu\nu} + \nabla_{\mu}\xi_{\nu} - \nabla_{\nu}\xi_{\mu}, \nonumber \\
C_{\mu} &\longrightarrow & C_{\mu} - \alpha\xi_{\mu},
\end{eqnarray}
and, $(ii)$ under the gauge transformation of St{\"u}ckelberg field $C_{\mu}$
\begin{eqnarray}
\label{slvb1}
C_{\mu} &\longrightarrow & C_{\mu} + \nabla_{\mu}\Lambda , \nonumber \\
B^{\mu\nu} &\longrightarrow & B^{\mu\nu} ,
\end{eqnarray}
where, $\xi_{\mu}$ and $\Lambda$ are the corresponding gauge parameters. 
In addition to the above symmetries of fields, there exists a set of transformation of gauge parameters $\Lambda$ and $\xi_{\mu}$ that leaves the fields $B_{\mu\nu}$ and $C_{\mu}$ invariant,
\begin{eqnarray}
\label{slvb2}
\xi_{\mu} &\longrightarrow & \xi_{\mu} + \nabla_{\mu}\psi , \nonumber \\
\Lambda &\longrightarrow & \Lambda + \alpha\psi ,
\end{eqnarray}
which means that the gauge generators are linearly dependent \cite{buchbinder2008}. The gauge fixing procedure for this Lagrangian is covered explicitly in Ref. \cite{aashish2018b}, and to save space we quote directly the final result for the total gauge fixed Lagrangian,
\begin{eqnarray}
\label{slvb3}
\mathcal{L}_{GF} = -\dfrac{1}{12}H_{\mu\nu\lambda}H^{\mu\nu\lambda} - \frac{1}{4}\alpha^{2}\Big(b_{\mu\nu}B^{\mu\nu}\Big)^{2} - \dfrac{1}{4}\Big(b_{\mu\nu}F^{\mu\nu}\Big)^{2} - \frac{1}{2}\Big(b_{\mu\nu}b_{\rho\sigma}\nabla^{\mu}B^{\rho\sigma}\Big)^{2} \nonumber \\ - \frac{1}{2}\alpha^{2}C_{\nu}C^{\nu} - \frac{1}{2}(\nabla_{\mu}\Phi)^{2} - \frac{1}{2}(\nabla^{\mu}C_{\mu})^{2} - \frac{1}{2}\alpha^{2}\Phi^{2}.
\end{eqnarray}
The presence of a new scalar field $\Phi$ is a direct consequence of gauge-fixing of St{\"u}ckelberg field, and explicitly displays a scalar degree of freedom that remains hidden in the original Lagrangian (\ref{slva5}) with broken gauge symmetry.

\section{\label{sec3}1-Loop Effective Action}
The quantization of theories such as (\ref{slva7}) is tricky, because of the symmetries in gauge parameters, as in Eq. (\ref{slvb2}). Such symmetries lead to a degeneracy in the ghost determinant appearing in the Faddeev-Popov procedure \cite{buchbinder1992}, and require special treatment for quantization \cite{buchbinder1988,buchbinder1992,aashish2018a}. We follow a general quantization procedure developed in Ref. \cite{aashish2018a} based on DeWitt-Vilkovisky's approach \cite{dewitt1967a,*dewitt1967b,*dewitt1967c,vilkovisky1984a} that yields covariant and background independent resutls, to deal with the additional symmetries of gauge parameters and derive the 1-loop effective action. 

For a quadratic action not involving quantization of metric the expression for 1-loop effective action in the DeWitt-Vilkovisky's field space notation, about a set of background fields $\bar{\varphi}^{i}$, is given by \cite{aashish2018a},
\begin{eqnarray}
\label{aea0}
\Gamma_{1}[\bar{\varphi}] = -\ln\det Q_{\alpha\beta}[\bar{\varphi}] + \ln\det \check{Q}_{a b} + \dfrac{1}{2}\ln\det\left(S^{GF}_{,ij}[\bar{\varphi}]]\right).
\end{eqnarray}
where $S^{GF}$ is the gauge-fixed action. Let us briefly explain the various (field-space) notations in Eq. (\ref{aea0}) (see \cite{parker2009} for a detailed introduction). The index $i$ in field space corresponds to all the tensor indices and spacetime dependence of fields in the coordinate space. For example, fields ($B_{\mu\nu}(x), C_{\mu}(x),\Phi(x)$) are denoted by components of $\varphi^{i}$ ($i = 1,2,3$) in field space, where $\varphi^{1}\leftrightarrow B_{\mu\nu}(x)$, $\varphi^{2}\leftrightarrow C_{\mu}(x)$, and $\varphi^{3}\leftrightarrow \Phi(x)$. The rest of the constructions in field space (tensors, scalar products, connections, field space metric, etc.) are similar to that in a coordinate space. The background fields in this notation, $\bar{\varphi}^{i}$, too carry all the indices of their respective counterparts including coordinate dependence. The object $S_{,ij}$ represents a derivative in field space, define by,
\begin{eqnarray}
\label{aea1}
S_{,ij}[\bar{\varphi}] = \left(\dfrac{\delta^2}{\delta \varphi^{j}\delta \varphi^{i}} S[\varphi]\right)_{\varphi = \bar{\varphi}}.
\end{eqnarray}
Let, $\check{\epsilon}^{a}$ parametrize the symmetry of gauge parameters, as in Eq. (\ref{slvb2}), and $\check{\chi}^{a}$ be the corresponding fixing condition for gauge parameters $\epsilon^{\alpha}$ (can be read off of Eqs. (\ref{slvb0}) and (\ref{slvb1}) ), then \cite{aashish2018a}
\begin{eqnarray}
\label{aea3}
Q^{\alpha}_{\beta} = \left(\dfrac{\delta}{\delta\epsilon^{\beta}}\chi^{\alpha}[\varphi,\epsilon,\check{\chi}]\right)_{\epsilon = 0},
\end{eqnarray}
where $\chi^{\alpha}$ is the gauge fixing condition for fields $\varphi^{i}$. $\det Q^{\alpha}_{\beta}$ is the ghost determinant factor. 

In the present case, corresponding to the symmetries (\ref{slvb0}), (\ref{slvb1}) and (\ref{slvb2}), there are two gauge conditions $\chi_{\xi_{\nu}} $ and $ \chi_{\Lambda}$, along with a condition $\check{\chi}_{\psi}$ on the parameters, that lead to three operators $Q^{\xi_{\mu}}_{\xi_{\nu}}$, $Q^{\Lambda}_{\Lambda}$ and $\check{Q}_{\psi}^{\psi}$ respectively. The results are displayed in TABLE \ref{tab1}.
\begin{table}[h!]
  \begin{center}
    \begin{tabular}{|c|c|} 
    \hline
      $\chi_{\xi_{\nu}} = b_{\mu\nu}b_{\rho\sigma}\nabla^{\mu} B^{\rho\sigma} + \alpha C_{\nu}$ & \ $Q^{\xi_{\mu}}_{\xi_{\nu}} = 2b_{\alpha\mu}b_{\beta\nu}\nabla^{\alpha}\nabla^{\beta} + \nabla_{\mu}\nabla_{\nu} - \alpha^{2}\delta_{\mu\nu}$ \ \\
      \hline
      $\chi_{\Lambda}= \nabla^{\mu}C_{\mu} + \alpha\Phi$ & $Q^{\Lambda}_{\Lambda} = \Box - \alpha^{2}$ \\
      \hline
      $\check{\chi}_{\psi}=\nabla^{\mu}\xi_{\mu} - \alpha\Lambda$ & $\check{Q}_{\psi}^{\psi}=\Box - \alpha^{2}$ \\
      \hline
    \end{tabular}
    \caption{Results for $Q$ operators corresponding to choices of fixing conditions $\chi$. }
    \label{tab1}
  \end{center}
\end{table}\\
Using these results in Eq. (\ref{aea0}), we get
\begin{eqnarray}
\label{aea4}
\Gamma_{1} = -\ln\det Q^{\xi_{\mu}}_{\xi_{\nu}} + \dfrac{1}{2}\ln\det\left(S^{GF}_{,ij}[\bar{\varphi}]]\right).
\end{eqnarray}
$S^{GF}$ is of course quadratic in fields, and the value of $\Gamma_{1}$ in operator form turns out to be, 
\begin{eqnarray}
\label{aea5}
\Gamma_{1} = \frac{i\hbar}{2}\Big[\ln\det(\Box_{2}{}^{\mu\nu}_{\ \ \rho\sigma} - \alpha^{2}b^{\mu\nu}b_{\rho\sigma}) - \ln\det(\Box_{1}{}^{\mu}_{\ \nu} -\alpha^{2}\delta^{\mu}_{\nu}) + \ln\det(\Box - \alpha^{2})\Big],
\end{eqnarray}
where,
\begin{eqnarray}
\label{aea6}
\Box_{2}{}^{\mu\nu}_{\ \ \rho\sigma}B^{\rho\sigma} &\equiv & \nabla_{\alpha}\nabla^{\alpha}B^{\mu\nu} + \nabla_{\alpha}\nabla^{\mu}B^{\nu\alpha} + \nabla_{\alpha}\nabla^{\nu}B^{\alpha\mu} + 2b^{\mu\nu}b_{\rho\sigma}b^{\alpha\sigma}b_{\beta\gamma}\nabla^{\rho}\nabla_{\alpha}B^{\beta\gamma}, \nonumber \\
\Box_{1}{}^{\mu}_{\ \nu}C^{\nu} &\equiv & 2b^{\nu\mu}b_{\rho\sigma}\nabla_{\nu}\nabla^{\rho}C^{\sigma} + \nabla^{\mu}\nabla_{\nu}C^{\nu};
\end{eqnarray}
and $\Box$ is the de'Alembertian operator.
In flat spacetime, no physically interesting inferences can be extracted from the above expression. However, in curved spacetime, the operators in Eq. (\ref{aea5}) are coupled to the metric $g_{\mu\nu}$. So, addressing certain issues, like that of quantum equivalence, then becomes nontrivial. Unfortunately, effective action cannot be calculated exactly in such cases \cite{aashish2018b}, and the best way forward is to perform a perturbative study. Therefore, we will consider a \textit{nearly} flat spacetime instead of a general curved one, so that,
\begin{eqnarray}
\label{aea7}
g_{\mu\nu}(x) = \eta_{\mu\nu} + \kappa h_{\mu\nu}(x).
\end{eqnarray}
$\eta_{\mu\nu}$ is the Minkowski metric and $h_{\mu\nu}$ is a perturbation, while $\kappa = 1/M_{p}$($M_{p}$ is Planck mass) parametrizes the scale of perturbation.

We can rewrite $\Gamma_{1}$ in integral form by introducing ghost fields $c_{\mu}$ and $\bar{c}_{\mu}$, 
\begin{eqnarray}
\label{aea8}
\Gamma_{1} = -\ln\int [d\eta][dc_{\mu}] [d\bar{c}_{\mu}] e^{-S_{GH}},
\end{eqnarray}
where, 
\begin{eqnarray}
\label{aea9}
S_{GH} = \eta^{i}S^{GF}_{,ij}\eta^{j} + \bar{c}^{\mu}Q^{\xi_{\mu}}_{\xi_{\nu}}c^{\nu},
\end{eqnarray}
and $\eta^{i}$ are the quantum fluctuations ($\delta B_{\mu\nu}(x), \delta C_{\mu}(x),\delta\Phi(x)$). Now, we use Eq. (\ref{aea7}) in Eq. (\ref{aea9}) and rearrange terms in orders of $h_{\mu\nu}$:
\begin{eqnarray}
\label{bea0}
S_{GH} = S_{0} + S_1 + O(h_{\mu\nu}h_{\alpha\beta}),
\end{eqnarray}
where the subscripts denote the power of $h_{\mu\nu}$. Substituting Eq. (\ref{bea0}) in Eq. (\ref{aea8}), and treating $S_{1}$ as a perturbation, the integrand can be Taylor expanded to write,
\begin{eqnarray}
\label{bea1}
\Gamma_{1} = -\ln\left(1 + \langle S_1\rangle + O(h_{\mu\nu}h_{\alpha\beta})\right),
\end{eqnarray}
where we have used the normalization for path integral of $S_0$. The logarithm can be further expanded to yield, upto first order in $h_{\mu\nu}$,
\begin{eqnarray}
\label{bea2}
\Gamma_1 = -\langle S_1\rangle .
\end{eqnarray}
The calculation of $\Gamma_1$ thus amounts to evaluating $\langle S_1\rangle$, which is a collection of two-point correlation functions of fields. These correlations are just the flat spacetime propagators of fields and can be derived from $S_0$ using projection operator method. We obtained the expansions of $S_{GH}$ using xAct packages \cite{xpert,xact} for Mathematica, results of which are presented below:
\begin{eqnarray}
\label{bea3}
S_0 &=& \int d^{4}x \Big(- \tfrac{1}{2} \alpha^2 \delta C_{\mu } \delta C^{\mu } -  \tfrac{1}{2} \alpha^2 \
\delta \phi^2 -  \tfrac{1}{4} \alpha^2 (\delta B^{\mu \nu } b_{\mu \nu })^2 \nonumber \\ 
&& -  \tfrac{1}{2} (b_{\alpha }{}^{\nu } b_{\beta \gamma } b_{\mu \nu \
} b_{\rho \sigma } \delta B^{\beta \gamma }{}^{,\alpha } \delta B^{\rho \sigma \
}{}^{,\mu }) -  \tfrac{1}{2} (\delta C_{\mu }{}^{,\mu })^2 \nonumber \\ 
&& -  \tfrac{1}{4} \bigl(b_{\mu \nu } (\delta C^{\nu }{}^{,\mu } -  \delta C^{\mu \
}{}^{,\nu })\bigr)^2 -  \tfrac{1}{2} \delta \phi {}_{,\mu } \delta \phi {}^{,\mu } \
\nonumber \\ 
&& + \delta \bar{c}^{\mu } (- \alpha^2 \delta c_{\mu } + \delta c^{\nu }{}_{,\nu }{}_{,\mu \
} + 2 b_{\nu \mu } b_{\rho \sigma } \delta c^{\sigma }{}^{,\rho }{}^{,\nu }) \
\nonumber \\ 
&& + \tfrac{1}{12} (- \delta B_{\nu \rho }{}_{,\mu } -  \delta B_{\rho \mu }{}_{,\nu } -  \delta B_{\mu \nu }{}_{,\rho }) (\delta B^{\nu \rho }{}^{,\mu } + \delta B^{\rho \
\mu }{}^{,\nu } + \delta B^{\mu \nu }{}^{,\rho }) \Big)
\end{eqnarray}
\begin{eqnarray}
\label{bea4}
S_1 &=& \int d^{4}x \Big( \tfrac{1}{2} \alpha^2 \delta C^{\mu } \delta C^{\nu } h{}_{\mu \nu } -  \alpha^2 \delta c^{\mu } \delta \bar{c}^{\nu } h{}_{\mu \nu } -  \tfrac{1}{4} \alpha^2 \delta \phi^2 h^{\mu }{}_{\mu } \nonumber \\ 
&& -  \tfrac{1}{4} \alpha^2 \delta C{}_{\mu } \delta C^{\mu } h^{\nu }{}_{\nu } -  \tfrac{1}{2} \alpha^2 \delta c^{\mu } \delta \bar{c}{}_{\mu } h^{\nu }{}_{\nu } + \tfrac{1}{2} \alpha^2 \delta B^{\mu \nu } \delta B^{\rho \sigma } h{}_{\rho a} b^{a}{}_{\sigma } n_{\mu \nu } \nonumber \\ 
&& + \tfrac{1}{2} \alpha^2 \delta B^{\mu \nu } \delta B^{\rho \sigma } h{}_{\sigma a} n_{\mu \nu } n_{\rho }{}^{a} -  \tfrac{1}{8} \alpha^2 \delta B^{\mu \nu } \delta B^{\rho \sigma } h^{a}{}_{a} n_{\mu \nu } n_{\rho \sigma } + \delta \bar{c}^{\mu } b^{\nu }{}_{\mu } b^{\rho \sigma } h{}_{\nu \rho }{}_{,a} \delta c{}_{\sigma }{}^{,a} \nonumber \\ 
&& - 2 \delta \bar{c}^{\mu } h{}_{\nu a} b^{\nu }{}_{\mu } b^{\rho \sigma } \delta c{}_{\sigma }{}_{,\rho }{}^{,a} + h{}_{\rho d} b^{bc} b^{\mu \nu } b^{\rho }{}_{\nu } b^{\sigma a} \delta B{}_{bc}{}^{,d} \delta B{}_{\sigma a}{}_{,\mu } + \tfrac{1}{2} h{}_{\rho a} b^{\mu \nu } b^{\rho \sigma } \delta C{}_{\sigma }{}^{,a} \delta C{}_{\nu }{}_{,\mu } \nonumber \\ 
&& + \tfrac{1}{2} \delta \bar{c}^{\mu } h^{\rho }{}_{\rho } \delta c^{\nu }{}_{,\nu }{}_{,\mu } -  \tfrac{1}{4} h^{\nu }{}_{\nu } \delta \phi{}_{,\mu } \delta \phi{}^{,\mu } + \tfrac{1}{2} h{}_{\sigma a} b^{\mu \nu } b^{\rho \sigma } \delta C{}_{\rho }{}^{,a} \delta C{}_{\mu }{}_{,\nu } \nonumber \\ 
&& -  \tfrac{1}{2} h{}_{\rho a} b^{\mu \nu } b^{\rho \sigma } \delta C{}_{\sigma }{}^{,a} \delta C{}_{\mu }{}_{,\nu } -  \tfrac{1}{4} h^{\rho }{}_{\rho } \delta C^{\mu }{}_{,\mu } \delta C^{\nu }{}_{,\nu } -  \tfrac{1}{2} \delta C^{\mu } h^{\rho }{}_{\rho }{}_{,\mu } \delta C^{\nu }{}_{,\nu } \nonumber \\ 
&& -  \tfrac{1}{2} \delta C^{\mu } b^{\nu \rho } b^{\sigma a} h{}_{\mu \sigma }{}_{,a} \delta C{}_{\rho }{}_{,\nu } + \delta \bar{c}^{\mu } b^{\nu }{}_{\mu } b^{\rho \sigma } h{}_{\rho \sigma }{}_{,a} \delta c^{a}{}_{,\nu } -  \delta \bar{c}^{\mu } b^{\nu }{}_{\mu } b^{\rho \sigma } \delta c{}_{\sigma }{}^{,a} h{}_{\rho a}{}_{,\nu } \nonumber \\ 
&& + \tfrac{1}{2} \delta \bar{c}^{\mu } \delta c^{\nu }{}_{,\mu } h^{\rho }{}_{\rho }{}_{,\nu } - 2 \delta \bar{c}^{\mu } h{}_{\rho a} b^{\nu }{}_{\mu } b^{\rho \sigma } \delta c{}_{\sigma }{}^{,a}{}_{,\nu } + \tfrac{1}{2} \delta c^{\mu } \delta \bar{c}^{\nu } h^{\rho }{}_{\rho }{}_{,\mu }{}_{,\nu } \nonumber \\ 
&& + \delta \bar{c}^{\mu } h^{a}{}_{a} b^{\nu }{}_{\mu } b^{\rho \sigma } \delta c{}_{\sigma }{}_{,\rho }{}_{,\nu } + \tfrac{1}{2} h{}_{\mu \nu } \delta \phi{}^{,\mu } \delta \phi{}^{,\nu } -  \tfrac{1}{4} h^{d}{}_{d} b^{bc} b^{\mu \nu } b^{\rho }{}_{\nu } b^{\sigma a} \delta B{}_{\sigma a}{}_{,\mu } \delta B{}_{bc}{}_{,\rho } \nonumber \\ 
&& + h{}_{cd} b^{bc} b^{\mu \nu } b^{\rho }{}_{\nu } b^{\sigma a} \delta B{}_{\sigma a}{}_{,\mu } \delta B{}_{b}{}^{d}{}_{,\rho } + h{}_{bd} b^{bc} b^{\mu \nu } b^{\rho }{}_{\nu } b^{\sigma a} \delta B{}_{\sigma a}{}_{,\mu } \delta B^{d}{}_{c}{}_{,\rho } + \tfrac{1}{2} h{}_{\sigma d} b^{ab} b^{cd} b^{\mu \nu } b^{\rho \sigma } \delta B{}_{ab}{}_{,c} \delta B{}_{\mu \nu }{}_{,\rho } \nonumber \\ 
&& + \tfrac{1}{2} h{}_{\sigma a} b^{\mu \nu } b^{\rho \sigma } \delta C{}_{\nu }{}_{,\mu } \delta C^{a}{}_{,\rho } + \tfrac{1}{2} \delta C^{\mu } b^{\nu \rho } b^{\sigma a} h{}_{\mu \sigma }{}_{,a} \delta C{}_{\nu }{}_{,\rho } -  \tfrac{1}{2} h{}_{\nu a} b^{\mu \nu } b^{\rho \sigma } \delta C{}_{\mu }{}^{,a} \delta C{}_{\sigma }{}_{,\rho } \nonumber \\ 
&& -  \tfrac{1}{8} h^{a}{}_{a} b^{\mu \nu } b^{\rho \sigma } \delta C{}_{\nu }{}_{,\mu } \delta C{}_{\sigma }{}_{,\rho } + \tfrac{1}{4} h^{a}{}_{a} b^{\mu \nu } b^{\rho \sigma } \delta C{}_{\mu }{}_{,\nu } \delta C{}_{\sigma }{}_{,\rho } + \delta \bar{c}^{\mu } b^{\nu }{}_{\mu } b^{\rho \sigma } h{}_{\nu \sigma }{}_{,a} \delta c^{a}{}_{,\rho } \nonumber \\ 
&& + \delta \bar{c}^{\mu } b^{\nu }{}_{\mu } b^{\rho \sigma } h{}_{\sigma a}{}_{,\nu } \delta c^{a}{}_{,\rho } + \delta C^{\mu } \delta C^{\nu }{}_{,\nu } h{}_{\mu }{}^{\rho }{}_{,\rho } + \tfrac{1}{4} \delta B^{\mu \nu } \delta B^{\rho \sigma }{}_{,\nu } h{}_{\mu \sigma }{}_{,\rho } \nonumber \\ 
&& -  \delta \bar{c}^{\mu } b^{\nu }{}_{\mu } b^{\rho \sigma } \delta c{}_{\sigma }{}^{,a} h{}_{\nu a}{}_{,\rho } + \delta \bar{c}^{\mu } b^{\nu }{}_{\mu } b^{\rho \sigma } \delta c^{a}{}_{,\nu } h{}_{\sigma a}{}_{,\rho } -  \delta c^{\mu } \delta \bar{c}^{\nu } b^{\rho }{}_{\nu } b^{\sigma a} h{}_{\mu \sigma }{}_{,a}{}_{,\rho } \nonumber \\ 
&& + \delta c^{\mu } \delta \bar{c}^{\nu } b^{\rho }{}_{\nu } b^{\sigma a} h{}_{\sigma a}{}_{,\mu }{}_{,\rho } + \delta c^{\mu } \delta \bar{c}^{\nu } b^{\rho }{}_{\nu } b^{\sigma a} h{}_{\mu a}{}_{,\sigma }{}_{,\rho } -  \tfrac{1}{8} g_{\mu \sigma } g_{\nu a} g_{\rho b} h^{c}{}_{c} \delta B^{\sigma a}{}^{,b} \delta B^{\mu \nu }{}^{,\rho } \nonumber \\ 
&& -  \tfrac{1}{4} g_{\mu a} g_{\nu b} g_{\rho \sigma } h^{c}{}_{c} \delta B^{\sigma a}{}^{,b} \delta B^{\mu \nu }{}^{,\rho } + \tfrac{1}{4} g_{\nu a} g_{\rho b} h{}_{\mu \sigma } \delta B^{\sigma a}{}^{,b} \delta B^{\mu \nu }{}^{,\rho } + \tfrac{1}{4} g_{\mu \sigma } g_{\rho b} h{}_{\nu a} \delta B^{\sigma a}{}^{,b} \delta B^{\mu \nu }{}^{,\rho } \nonumber \\ 
&& + \tfrac{1}{2} g_{\mu a} g_{\rho \sigma } h{}_{\nu b} \delta B^{\sigma a}{}^{,b} \delta B^{\mu \nu }{}^{,\rho } + \tfrac{1}{2} g_{\mu b} g_{\rho a} h{}_{\nu \sigma } \delta B^{\sigma a}{}^{,b} \delta B^{\mu \nu }{}^{,\rho } + \tfrac{1}{4} g_{\mu \sigma } g_{\nu a} h{}_{\rho b} \delta B^{\sigma a}{}^{,b} \delta B^{\mu \nu }{}^{,\rho } \nonumber \\ 
&& + \tfrac{1}{2} g_{\mu a} g_{\nu b} h{}_{\rho \sigma } \delta B^{\sigma a}{}^{,b} \delta B^{\mu \nu }{}^{,\rho } + h{}_{\nu \rho } \delta C^{\mu }{}_{,\mu } \delta C^{\nu }{}^{,\rho } + \tfrac{1}{2} \delta B^{\mu \nu } b^{b}{}_{a} b^{cd} b^{\rho }{}_{\nu } b^{\sigma a} h{}_{\mu \rho }{}_{,b} \delta B{}_{cd}{}_{,\sigma } \nonumber \\ 
&& + \tfrac{1}{2} \delta B^{\mu \nu } b^{b}{}_{a} b^{cd} n_{\mu }{}^{\rho } b^{\sigma a} h{}_{\nu \rho }{}_{,b} \delta B{}_{cd}{}_{,\sigma } -  \tfrac{1}{2} \delta B^{\mu \nu } b^{b}{}_{a} b^{cd} b^{\rho }{}_{\nu } b^{\sigma a} h{}_{\rho b}{}_{,\mu } \delta B{}_{cd}{}_{,\sigma } -  \tfrac{1}{2} \delta B^{\mu \nu } b^{b}{}_{a} b^{cd} n_{\mu }{}^{\rho } b^{\sigma a} h{}_{\rho b}{}_{,\nu } \delta B{}_{cd}{}_{,\sigma } \nonumber \\ 
&& + \tfrac{1}{2} \delta B^{\mu \nu } b^{b}{}_{a} b^{cd} b^{\rho }{}_{\nu } b^{\sigma a} h{}_{\mu b}{}_{,\rho } \delta B{}_{cd}{}_{,\sigma } + \tfrac{1}{2} \delta B^{\mu \nu } b^{b}{}_{a} b^{cd} n_{\mu }{}^{\rho } b^{\sigma a} h{}_{\nu b}{}_{,\rho } \delta B{}_{cd}{}_{,\sigma } -  \tfrac{1}{2} h{}_{\rho a} b^{\mu \nu } b^{\rho \sigma } \delta C{}_{\nu }{}_{,\mu } \delta C^{a}{}_{,\sigma } \nonumber \\ 
&& -  \tfrac{1}{2} h{}_{\nu a} b^{\mu \nu } b^{\rho \sigma } \delta C^{a}{}_{,\mu } \delta C{}_{\rho }{}_{,\sigma } + \tfrac{1}{2} h{}_{\mu a} b^{\mu \nu } b^{\rho \sigma } \delta C^{a}{}_{,\nu } \delta C{}_{\rho }{}_{,\sigma } -  \tfrac{1}{8} h^{a}{}_{a} b^{\mu \nu } b^{\rho \sigma } \delta C{}_{\mu }{}_{,\nu } \delta C{}_{\rho }{}_{,\sigma } \nonumber \\ 
&& + \tfrac{1}{2} \delta C^{\mu } b^{\nu \rho } b^{\sigma a} \delta C{}_{\rho }{}_{,\nu } h{}_{\mu a}{}_{,\sigma } -  \tfrac{1}{2} \delta C^{\mu } b^{\nu \rho } b^{\sigma a} \delta C{}_{\nu }{}_{,\rho } h{}_{\mu a}{}_{,\sigma } -  \delta \bar{c}^{\mu } b^{\nu }{}_{\mu } b^{\rho \sigma } \delta c^{a}{}_{,\rho } h{}_{\nu a}{}_{,\sigma } \nonumber \\ 
&& + \tfrac{1}{4} \delta B^{\mu \nu } \delta B^{\rho \sigma }{}_{,\mu } h{}_{\nu \rho }{}_{,\sigma } -  \delta \bar{c}^{\mu } b^{\nu }{}_{\mu } b^{\rho \sigma } \delta c^{a}{}_{,\nu } h{}_{\rho a}{}_{,\sigma } + \tfrac{1}{4} \delta B^{\mu \nu } h{}_{\nu \rho }{}_{,\sigma } \delta B{}_{\mu }{}^{\rho }{}^{,\sigma } \nonumber \\ 
&& + \tfrac{1}{4} \delta B^{\mu \nu } h{}_{\mu \sigma }{}_{,\rho } \delta B{}_{\nu }{}^{\rho }{}^{,\sigma } + \tfrac{1}{4} \delta B^{\mu \nu } h{}_{\nu \sigma }{}_{,\rho } \delta B^{\rho }{}_{\mu }{}^{,\sigma } + \tfrac{1}{4} \delta B^{\mu \nu } h{}_{\mu \rho }{}_{,\sigma } \delta B^{\rho }{}_{\nu }{}^{,\sigma } \Big)
\end{eqnarray}

\subsection{Propagators}
We use the projection operator method \cite{barnes1965} to invert the operators in $S_0$ and derive the Green's functions or propagators. In the operator form, $S_0$ can be recast as
\begin{eqnarray}
\label{apr0}
S_0 =\int d^{4}x \Big(  \dfrac{1}{4} B^{\mu\nu}\mathcal{O}^{B}_{\mu\nu,\alpha\beta}B^{\alpha\beta} + \dfrac{1}{2} C^{\mu}\mathcal{O}^{C}_{\mu\nu}C^{\nu} + \frac{1}{2}\Phi\mathcal{O}^{\Phi}\Phi\Big)
\end{eqnarray}
where,
\begin{eqnarray}
\label{apr1}
\mathcal{O}^{B}_{\mu\nu,\alpha\beta} &=& \frac{\Box}{2}(\eta_{\mu\alpha}\eta_{\nu\beta}-\eta_{\mu\beta}\eta_{\nu\alpha})  + \frac{1}{2}(\partial_{\mu}\partial_{\beta}\eta_{\nu\alpha}\!-\!\partial_{\nu}\partial_{\beta}\eta_{\mu\alpha}\!-\!\partial_{\mu}\partial_{\alpha}\eta_{\nu\beta}\!+\!\partial_{\nu}\partial_{\alpha}\eta_{\mu\beta})\nonumber\\ && -\,\big(\alpha^2 + 2 (b_{\rho\sigma}\partial^{\rho})^2\big) b_{\mu\nu}b_{\alpha\beta},\\
\label{apr2}
\mathcal{O}^{C}_{\mu\nu} &=& 2 b_{\sigma\mu}b_{\rho\nu}\partial^{\sigma}\partial^{\rho} + \partial_{\mu}\partial_{\nu} - \alpha^{2}\eta_{\mu\nu},\\
\label{apr3}
\mathcal{O}^{\Phi} &=& \Box - \alpha^{2}.
\end{eqnarray}
At this point, we would like to point out that a calculation for the propagator of $B_{\mu\nu}$ using projector method was first performed in Ref. \cite{maluf2019} recently. However, their calculation did not account for the St\"uckelberg field and as a result our operator $(\mathcal{O}^{B})_{\mu\nu,\alpha\beta}$ is different from the one in Ref. \cite{maluf2019}, which misses the contribution from gauge-fixing term  $2 (b_{\rho\sigma}\partial^{\rho})^2 b_{\mu\nu}b_{\alpha\beta}$. Fortunately, this term is merely an addition to mass, $\alpha^{2}$, and ends up not contributing to the propagator, $(\mathcal{O}^{B})^{-1}_{\mu\nu,\alpha\beta}$. So, we end up getting an identical result for the propagator, barring complex infinity terms that can be ignored (see appendix for details of projection operators $P^{(1)},...,P^{(6)}$),
\begin{eqnarray}
\label{apr4}
(\mathcal{O}^{B})^{-1}_{\mu\nu,\alpha\beta} (x,x')=\int \dfrac{d^{4}p}{(2\pi^{n})}e^{-ip\cdot(x-x')}\left(\frac{1}{p^{2}} P^{(1)}_{\mu\nu, \alpha\beta} + \frac{b^{2}}{(b_{\rho\sigma}p^{\sigma})^{2}}( P^{(4)}_{\mu\nu, \alpha\beta}+ P^{(5)}_{\mu\nu, \alpha\beta})\right),
\end{eqnarray}
There are no massive propagating modes in Eq. (31) and only one massless mode propagates, as concluded in Ref. \cite{altschul2010,maluf2019}. The second pole describes a massless pole propagating in an anisotropic medium, which for our choice of $b_{\mu\nu}$ gives,
\begin{eqnarray}
\label{bpr0}
b^{2}\big((p^{2})^{2} + (p^{3})^{2}\big) = 0.
\end{eqnarray}
Contrary to the claim in Ref. \cite{maluf2019} where these modes were described as non-physical due to a negative sign appearing in energy-momentum relations as a result of a different choice of $b_{\mu\nu}$, we note that for our choice of $b_{\mu\nu}$ which corresponds to monopole solutions of antisymmetric tensor, energy terms ($p^0$) disappear altogether. 

For the St\"uckelberg field $C_{\mu}$, spontaneous Lorentz violating term appears in the kinetic part (first term in Eq. \ref{apr2}), which makes inverting $\mathcal{O}^{C}_{\mu\nu}$ a little tricky. New projector operators have to be defined apart from the longitudinal and transverse momentum operators, that also have a closed algebra, so that any operator $D_{\mu\nu}$ can be then expanded in terms of these projectors. We define,
\begin{eqnarray}
\label{apr5}
\mathcal{P}^{(1)}_{\mu\nu} = \dfrac{p_{\mu}p_{\nu}}{p^2}; \quad \mathcal{P}^{(2)}_{\mu\nu} = \eta_{\mu\nu} - \dfrac{p_{\mu}p_{\nu}}{p^2}; \nonumber \\
\mathcal{P}^{(3)}_{\mu\nu} = \dfrac{1}{(b_{\rho\sigma}p^{\sigma})^{2}} b_{\sigma\mu}b_{\rho\nu}p^{\sigma}p^{\rho} .
\end{eqnarray}
These operators satisfy a closed algebra, as shown in TABLE \ref{tab2}.
\begin{table}[h!]
  \begin{center}
    \begin{tabular}{|c|c|c|c|} 
    \hline
      & $\mathcal{P}^{(1)}$ & $\mathcal{P}^{(2)}$ & $\mathcal{P}^{(3)}$ \\
      \hline
    $\mathcal{P}^{(1)}$  & $\mathcal{P}^{(1)}$ & 0 & 0 \\
      \hline
    $ \mathcal{P}^{(2)}$ & 0 & $ \mathcal{P}^{(2)}$ & $\mathcal{P}^{(3)}$ \\
      \hline
     $\mathcal{P}^{(3)}$ & 0 & $\mathcal{P}^{(3)}$ & $\mathcal{P}^{(3)}$ \\
      \hline
    \end{tabular}
    \caption{Algebra of projection operators for the St\"uckelberg field $C_{\mu}$. Tensor indices have not been explicitly written. }
    \label{tab2}
  \end{center}
\end{table}
Using these operators, $\mathcal{O}^{C}_{\mu\nu}$ in momentum space can be written as,
\begin{eqnarray}
\label{apr6}
\mathcal{O}^{C}_{\mu\nu} = - 2(b_{\rho\sigma}p^{\sigma})^2 \mathcal{P}^{(3)}_{\mu\nu} - \left(p^2 + \alpha^2 \right) \mathcal{P}^{(1)}_{\mu\nu} - \alpha^2 \mathcal{P}^{(2)}_{\mu\nu}.
\end{eqnarray}
Assuming that $(\mathcal{O}^{C})^{-1}_{\mu\nu}$ in momentum space has the form,
\begin{eqnarray}
\label{apr7}
(\mathcal{O}^{C})^{-1}_{\mu\nu} = m_{1}\mathcal{P}^{(1)}_{\mu\nu} + m_{2}\mathcal{P}^{(2)}_{\mu\nu} + m_{3} \mathcal{P}^{(3)}_{\mu\nu},
\end{eqnarray}
we use the identity $\mathcal{O}\mathcal{O}^{-1} = \mathcal{I}$ to obtain,
\begin{eqnarray}
\label{apr8}
(\mathcal{O}^{C})^{-1}_{\mu\nu}(x,x') = \int \dfrac{d^{4}p}{(2\pi^{n})}e^{-ip\cdot(x-x')}\left(-\dfrac{1}{p^2 + \alpha^2}\mathcal{P}^{(1)}_{\mu\nu} - \dfrac{1}{\alpha^{2}}\mathcal{P}^{(2)}_{\mu\nu} + \dfrac{1}{\alpha^2}\dfrac{(b_{\rho\sigma}p^{\sigma})^{2}}{(b_{\rho\sigma}p^{\sigma})^{2} + \alpha^2/2} \mathcal{P}^{(3)}_{\mu\nu}\right).
\end{eqnarray}
Here, a massive scalar mode with pole at $\alpha$ propagates while another anisotropic mode propagates with mass $\alpha/\sqrt{2}$. Terms in $\mathcal{P}^{(2)}$ contain a massless pole and an additive poleless term which does not contribute to correlations and can be ignored. For $\Phi$, the scalar propagator is given by,
\begin{eqnarray}
\label{bpr1}
(\mathcal{O}^{\Phi})^{-1} (x,x') = \int \dfrac{d^{4}p}{(2\pi^{n})}e^{-ip\cdot(x-x')}\dfrac{1}{p^{2} + \alpha^2}.
\end{eqnarray}

\subsection{Quantum corrections}
Since all terms in $\langle S_{1}\rangle$ are local, they correspond to tadpole diagrams. We solve these integrals in two steps: first, the derivatives of field fluctuations are transformed to momentum space by substituting Eqs. (\ref{apr4}), (\ref{apr8}), and (\ref{bpr1}). We also perform by-parts integrals to get rid of derivatives of $h_{\mu\nu}$, so that in all expressions below, a coefficient $h_{\mu\nu}$ is understood to be present but not explicitly written. The Fourier transformed $\langle S_{1}\rangle$ then has terms of the form,
\begin{eqnarray}
\label{aqc0}
\int d^{4}x A(x) \langle\partial^{m}\delta \ \partial^{n}\delta\rangle \longrightarrow \int d^{4}x \dfrac{d^{4}p}{(2\pi^{n})} A(x) (-ip)^{m} (ip)^{n} \langle \delta_{p}\delta_{p}\rangle,
\end{eqnarray}
where, tensor indices of $A(x)$ and $\delta$ have been omitted for convenience. $\delta$ is the quantum field fluctuation, and $\langle \delta_{p}\delta_{p}\rangle$ represents the propagator(s) in momentum space. 

The second step is to replace $\langle \delta_{p}\delta_{p}\rangle$ with values of Green's function and evaluate the integrals. We primarily use the results in Ref. \cite{bardin1999} to evaluate the divergent terms of most of the integrals, except those involving anisotropic term $(b_{\rho\sigma}p^{\sigma})^{2}$. There are two types of poleless integrals coming from Eq. (\ref{apr4}):
\begin{eqnarray}
\label{aqc1}
\int d^{4}x \dfrac{d^{4}p}{(2\pi^{n})} A(x)\dfrac{p^{\mu}...p^{\beta}}{p^{2}} ; \quad \int d^{4}x \dfrac{d^{4}p}{(2\pi^{n})} A(x)\dfrac{p^{\mu}...p^{\beta}}{(b_{\rho\sigma}p^{\sigma})^{2}},
\end{eqnarray}
with upto four $p^{\mu}$'s in the numerator. The first integral vanishes due to the lack of a physical scale \cite{bardin1999}. To solve the second integral, we use the approach developed in \cite{george1975,*george1987,capper1982a,*capper1982b}, and find that it also does not have any physical contribution. 

Next, there are broadly three types of integrals with non-zero poles arising from the rest of propagators:
\begin{eqnarray}
\label{aqc2}
&\int d^{4}x \dfrac{d^{4}p}{(2\pi^{n})} A(x)\dfrac{p^{\mu}...p^{\beta}}{p^{2} + \alpha^2} ; \quad \int d^{4}x \dfrac{d^{4}p}{(2\pi^{n})} A(x)\dfrac{p^{\mu}...p^{\beta}}{p^2(p^{2} + \alpha^2)};& \nonumber \\
 &\int d^{4}x \dfrac{d^{4}p}{(2\pi^{n})} A(x)\dfrac{p^{\mu}...p^{\beta}}{(b_{\rho\sigma}p^{\sigma})^{2} + \alpha^2/2};&
\end{eqnarray}
Again, the solutions to first two types of integrals are available in Ref. \cite{bardin1999}. We solve the third type of integral as follows. Following \cite{george1975}, we write
\begin{eqnarray}
\label{aqc3}
\int d^{4}p \dfrac{1}{(b_{\rho\sigma}p^{\sigma})^{2} + \alpha^2/2} = \int d^{4}p \int_{0}^{\infty} d\theta \exp[-\theta((b_{\rho\sigma}p^{\sigma})^{2} + \alpha^2/2)].
\end{eqnarray}
Integrating over $d^{4}p$, followed by writing the integral over $\theta$ in terms of $\Gamma$ function leads to familiar expressions encountered in dimensional regularization, which finally yields the divergent part as ($\epsilon = n-4$),
\begin{eqnarray}
\label{aqc4}
divp\left(\int d^{4}p \dfrac{1}{(b_{\rho\sigma}p^{\sigma})^{2} + \alpha^2/2}\right) = -\dfrac{\pi^{2}\alpha^{2}}{2\sqrt{\det(b_{\mu\rho}b^{\rho\nu})}}\dfrac{2}{\epsilon},
\end{eqnarray} 
which is identical to that of a scalar propagator integral except for the $\sqrt{\det(b_{\mu\rho}b^{\rho\nu})}$ in the denominator. For our choice of $b_{\mu\nu}$, Eq. (\ref{slva1}) with $a=0$ and $b=1/\sqrt{2}$, this term becomes a diagonal matrix,
\begin{eqnarray}
\label{aqc5}
b_{\mu\rho}b^{\rho\nu} = diag(0 \ \ 0 \ \ 1/2 \ \ 1/2),
\end{eqnarray}
implying that the determinant is zero. It turns out however, that this determinant appears as a factor in the denominator of the divergent part of effective action, and hence we use a regularization factor $\epsilon'$ to write,
 \begin{eqnarray}
\label{aqc6}
b_{\mu\rho}b^{\rho\nu} = \lim_{\epsilon'\to 0} diag(\epsilon' \ \ \epsilon' \ \ 1/2 \ \ 1/2).
\end{eqnarray}

With these inputs in xAct\cite{xact}, the final result for the divergent part of 1-loop effective after some further manipulations, is obtained as,
\begin{eqnarray}
\label{1lea}
divp(\Gamma_{1}) &=& \dfrac{1}{16\pi^{2}\epsilon}\Big( \alpha^4 \kappa h^{a }{}_{a }  + \dfrac{1}{\sqrt{\det(b_{\mu\rho}b^{\rho\nu})}} \Big( -  \tfrac{1}{16} \alpha^4 \kappa h{}_{\mu b} b_{a}{}^{b} b^{\mu a} -  \tfrac{3}{32} \alpha^4 \kappa h^{b}{}_{b} b_{a\mu } b^{\mu a} \nonumber \\ 
&& -  \tfrac{1}{12} \alpha^4 \kappa h{}_{\mu c} b_{a}{}^{b} b_{b}{}^{\nu } b^{c}{}_{\nu } b^{\mu a} -  \tfrac{5}{192} \alpha^4 \kappa h{}_{bc} b_{a\mu } b^{b\nu } b^{c}{}_{\nu } b^{\mu a} + \tfrac{1}{192} \alpha^4 \kappa h{}_{\nu c} b_{b}{}^{c} b^{b\nu } b_{\mu a} b^{\mu a} \nonumber \\ 
&& + \tfrac{1}{96} \alpha^4 \kappa h{}_{bc} b^{b\nu } b^{c}{}_{\nu } b_{\mu a} b^{\mu a} -  \tfrac{1}{384} \alpha^4 \kappa h^{c}{}_{c} b_{a}{}^{\nu } b_{b\nu } b_{\mu }{}^{b} b^{\mu a} -  \tfrac{1}{48} \alpha^4 \kappa h{}_{\nu c} b_{a}{}^{\nu } b^{c}{}_{b} b_{\mu }{}^{b} b^{\mu a} \nonumber \\ 
&& + \tfrac{5}{192} \alpha^4 \kappa h{}_{bc} b_{a}{}^{\nu } b^{c}{}_{\nu } b_{\mu }{}^{b} b^{\mu a} -  \tfrac{1}{384} \alpha^4 \kappa h^{c}{}_{c} b_{a\mu } b^{b\nu } b^{\mu a} b_{\nu b} + \tfrac{5}{384} \alpha^4 \kappa h^{c}{}_{c} b^{b\nu } b_{\mu a} b^{\mu a} b_{\nu b} \nonumber \\ 
&& + \tfrac{5}{192} \alpha^4 \kappa h^{c}{}_{c} b_{a}{}^{\nu } b_{\mu }{}^{b} b^{\mu a} b_{\nu b} + \tfrac{1}{192} \alpha^4 \kappa h{}_{\mu c} b_{a}{}^{b} b_{b}{}^{\nu } b^{\mu a} b_{\nu }{}^{c} + \tfrac{1}{192} \alpha^4 \kappa h{}_{bc} b_{a\mu } b^{b\nu } b^{\mu a} b_{\nu }{}^{c} \nonumber \\ 
&& -  \tfrac{7}{192} \alpha^4 \kappa h{}_{bc} b^{b\nu } b_{\mu a} b^{\mu a} b_{\nu }{}^{c} -  \tfrac{1}{96} \alpha^4 \kappa h{}_{bc} b_{a}{}^{\nu } b_{\mu }{}^{b} b^{\mu a} b_{\nu }{}^{c} -  \tfrac{1}{384} \alpha^4 \kappa h^{c}{}_{c} b_{a}{}^{b} b_{b}{}^{\nu } b^{\mu a} b_{\nu \mu } \nonumber \\ 
&& -  \tfrac{1}{96} \alpha^4 \kappa h{}_{bc} b^{c}{}_{\nu } b_{\mu }{}^{b} b^{\mu a} b^{\nu }{}_{a} + \tfrac{1}{96} \alpha^4 \kappa h{}_{\mu c} b_{a}{}^{b} b^{c}{}_{\nu } b^{\mu a} b^{\nu }{}_{b}\Big)\Big)
\end{eqnarray}
where $\epsilon = n-4$ (as $n\to 4$) is the divergence parameter from dimensional regularization. Eq. (\ref{1lea}) presents the divergent piece of one loop corrections of antisymmetric tensor field theory with spontaneous Lorentz violation at leading order in field fluctuations in a nearly flat spacetime, and is valid for a vacuum value that supports monopole solutions. The one-loop divergence structures in principle lead to corrections to parameters (or couplings) in the classical action through counterterms (for example, in Ref. \cite{mackay2010}). Studying such corrections is interesting at higher orders in background fields, but lie beyond the scope of present work. Also, it is not easy to compare theories with and without Lorentz violation in the present context, because the simplest Lorentz violating potential contains upto quartic order terms in fields; while without Lorentz violation, the potential(s) that have been studied in the past \cite{buchbinder2008} are quadratic in field components.

\section{\label{sec4}Quantum Equivalence}
The classical Lagrangian (\ref{slva5}) can be written in an equivalent form where the field $B_{\mu\nu}$ can be eliminated through the introduction of a vector field, so that the resulting Lagrangian describes a classically equivalent vector theory with spontaneous Lorentz violation. In this section, we will check their quantum equivalence at one-loop level. 

Checking classical equivalence of two theories is an interesting theoretical exercise, because it provides insight into the degrees of freedom and dynamical properties of theories that may be described by very different fields, like in 2-form, 1-form or a scalar field theories, and thus may lead to several simplifications in a given theory. This problem naturally extends to the quantum regime, and it is certainly not trivial to prove quantum equivalence of two classically equivalent theories especially in curved spacetime. For instance, it can be shown that a massive 2-form field is quantum equivalent to a massive vector field because of some special topological properties of zeta functions \cite{buchbinder2008}. However, it is extremely difficult to perform similar analyses when, for example, the Lorentz symmetry is broken \cite{aashish2018b}. In flat spacetime, establishing quantum equivalence is indeed trivial, because there is no field dependence in $\Gamma_{1}$ (Eq. (\ref{aea5})) and hence effective actions of two theories do not possess any physical distinction.

On the contrary, in curved spacetime, the presence of metric makes things interesting. Only problem is, the effective action cannot be calculated exactly. So, our best bet, in this case, is to do a perturbative study like the one in the previous section. 

Classical equivalence of Eq. (\ref{slva5}) was explored in Ref. \cite{altschul2010}, it was found to be equivalent to,
\begin{eqnarray}
\label{eq0}
\mathcal{L} = \dfrac{1}{2}B_{\mu\nu}\mathcal{F}^{\mu\nu} - \frac{1}{2}C^{\mu}C_{\mu} - \dfrac{1}{4}\alpha^{2} \Big(b_{\mu\nu}B^{\mu\nu}\Big)^{2},
\end{eqnarray}
where,
\begin{eqnarray}
\label{eq1}
\mathcal{F}_{\mu\nu} = \dfrac{1}{2}\epsilon_{\mu\nu\rho\sigma}F^{\mu\nu}.
\end{eqnarray}
$C_{\mu}$ is a vector field and $F_{\mu\nu}$ is as defined before. We choose to continue with the same symbol for vector and St\"uckelberg field to avoid unnecessary complications. Eq. (\ref{eq0}) can be written exclusively in terms of $C_{\mu}$ through the use of projection operators,
\begin{eqnarray}
\label{eq2}
T_{||\mu\nu} = b_{\rho\sigma}T^{\rho\sigma} b_{\mu\nu}, \nonumber \\
T_{\perp\mu\nu} = T_{\mu\nu} - T_{||\mu\nu},
\end{eqnarray}
for any two-rank tensor $T_{\mu\nu}$, and subsequently using the equations of motion for $B_{||\mu\nu}$ and $B_{\perp\mu\nu}$, to obtain,
\begin{eqnarray}
\label{eq3}
\alpha^{2}\mathcal{L} = \dfrac{1}{4}\left(\tilde{b}_{\mu\nu}F^{\mu\nu}\right)^{2} - \frac{1}{2}\alpha^{2} C^{\mu}C_{\mu},
\end{eqnarray}
where we have defined $\tilde{b}_{\mu\nu} = \dfrac{1}{2}\epsilon_{\mu\nu\rho\sigma}b^{\rho\sigma}$. Note that Lorentz violation enters Eq. (\ref{eq3}) through the kinetic term, although it is still gauge-symmetric. A similar exercise of applying St\"uckelberg procedure leads to the gauge fixed action in flat spacetime,
\begin{eqnarray}
\label{eq4}
\tilde{S}_{0} = \int d^{4}x \Big( \dfrac{1}{2} C^{\mu}\mathcal{O}^{C'}_{\mu\nu}C^{\nu} + \frac{1}{2}\Phi\mathcal{O}^{\Phi}\Phi \Big)
\end{eqnarray}
where, 
\begin{eqnarray}
\label{eq5}
\mathcal{O}^{C'}_{\mu\nu} = -\dfrac{1}{2}\big(\tilde{b}_{\sigma\mu}\tilde{b}_{\rho\nu}\partial^{\sigma}\partial^{\rho} + \tilde{b}_{\sigma\nu}\tilde{b}_{\rho\mu}\partial^{\sigma}\partial^{\rho}\big) + \dfrac{1}{2}\partial_{\mu}\partial_{\nu} - \dfrac{1}{2}\alpha^{2}\eta_{\mu\nu}.
\end{eqnarray}
A similar calculation of the propagator yields,
\begin{eqnarray}
\label{eq6}
(\mathcal{O}^{C'})^{-1}_{\mu\nu}(x,x') = \int \dfrac{d^{4}p}{(2\pi^{n})}e^{-ip\cdot(x-x')}\left(-\dfrac{1}{p^2 + \alpha^2}\mathcal{P}^{(1)}_{\mu\nu} - \dfrac{1}{\alpha^{2}}\mathcal{P}^{(2)}_{\mu\nu} - \dfrac{1}{\alpha^2}\dfrac{(\tilde{b}_{\rho\sigma}p^{\sigma})^{2}}{-(\tilde{b}_{\rho\sigma}p^{\sigma})^{2} + \alpha^2/2} \tilde{\mathcal{P}}^{(3)}_{\mu\nu}\right),
\end{eqnarray}
where $\tilde{\mathcal{P}}^{(3)}_{\mu\nu}$ has the same form as $\mathcal{P}^{3}_{\mu\nu}$ but with $\tilde{b}_{\mu\nu}$ instead of $b_{\mu\nu}$. Finally, the one-loop effective action is found to be,
\begin{eqnarray}
\label{eq7}
divp(\tilde{\Gamma}_{1}) &=&  \dfrac{1}{16\pi^{2}\epsilon}\Big( \alpha^4 \kappa h^{a }{}_{a }  + \dfrac{1}{\sqrt{-\det(\tilde{b}_{\mu\rho}\tilde{b}^{\rho\nu})}} \Big(\tfrac{1}{96} \alpha^4 \kappa h{}_{ec} \tilde{b}_{ad} \tilde{b}^{ad} \tilde{b}_{b}{}^{c} \tilde{b}^{be} + \tfrac{1}{32} \alpha^4 \kappa h^{b}{}_{b} \tilde{b}^{ac} \tilde{b}_{ca} \nonumber \\ 
&& -  \tfrac{1}{16} \alpha^4 \kappa h{}_{ab} \tilde{b}^{ac} \tilde{b}_{c}{}^{b} + \tfrac{1}{192} \alpha^4 \kappa h{}_{bc} \tilde{b}_{ad} \tilde{b}^{ad} \tilde{b}^{be} \tilde{b}^{c}{}_{e} -  \tfrac{1}{96} \alpha^4 \kappa h{}_{ac} \tilde{b}^{ad} \tilde{b}_{b}{}^{e} \tilde{b}^{c}{}_{e} \tilde{b}_{d}{}^{b} \nonumber \\ 
&& + \tfrac{1}{96} \alpha^4 \kappa h{}_{ec} \tilde{b}_{a}{}^{b} \tilde{b}^{ad} \tilde{b}_{b}{}^{c} \tilde{b}_{d}{}^{e} -  \tfrac{1}{384} \alpha^4 \kappa h^{c}{}_{c} \tilde{b}_{a}{}^{b} \tilde{b}^{ad} \tilde{b}_{be} \tilde{b}_{d}{}^{e} -  \tfrac{1}{96} \alpha^4 \kappa h{}_{ec} \tilde{b}_{a}{}^{b} \tilde{b}^{ad} \tilde{b}^{c}{}_{b} \tilde{b}_{d}{}^{e} \nonumber \\ 
&& + \tfrac{5}{192} \alpha^4 \kappa h{}_{bc} \tilde{b}_{a}{}^{b} \tilde{b}^{ad} \tilde{b}^{c}{}_{e} \tilde{b}_{d}{}^{e} -  \tfrac{1}{384} \alpha^4 \kappa h^{c}{}_{c} \tilde{b}^{ad} \tilde{b}_{b}{}^{e} \tilde{b}_{d}{}^{b} \tilde{b}_{ea} + \tfrac{1}{384} \alpha^4 \kappa h^{c}{}_{c} \tilde{b}_{ad} \tilde{b}^{ad} \tilde{b}^{be} \tilde{b}_{eb} \nonumber \\ 
&& -  \tfrac{1}{384} \alpha^4 \kappa h^{c}{}_{c} \tilde{b}^{ad} \tilde{b}^{be} \tilde{b}_{da} \tilde{b}_{eb} + \tfrac{1}{192} \alpha^4 \kappa h^{c}{}_{c} \tilde{b}_{a}{}^{b} \tilde{b}^{ad} \tilde{b}_{d}{}^{e} \tilde{b}_{eb} -  \tfrac{1}{64} \alpha^4 \kappa h{}_{bc} \tilde{b}_{ad} \tilde{b}^{ad} \tilde{b}^{be} \tilde{b}_{e}{}^{c} \nonumber \\ 
&& + \tfrac{1}{192} \alpha^4 \kappa h{}_{bc} \tilde{b}^{ad} \tilde{b}^{be} \tilde{b}_{da} \tilde{b}_{e}{}^{c} + \tfrac{1}{192} \alpha^4 \kappa h{}_{ac} \tilde{b}^{ad} \tilde{b}_{b}{}^{e} \tilde{b}_{d}{}^{b} \tilde{b}_{e}{}^{c} -  \tfrac{1}{48} \alpha^4 \kappa h{}_{bc} \tilde{b}_{a}{}^{b} \tilde{b}^{ad} \tilde{b}_{d}{}^{e} \tilde{b}_{e}{}^{c} \nonumber \\ 
&& -  \tfrac{1}{192} \alpha^4 \kappa h{}_{db} \tilde{b}_{a}{}^{b} \tilde{b}^{ad} \tilde{b}_{ce} \tilde{b}^{ec}\Big)\Big)
\end{eqnarray}
Upon comparing Eqs. (\ref{1lea}) and (\ref{eq7}), we can immediately notice that the first term is identical, while the rest of terms appearing with $b_{\mu\nu}$ and $\tilde{b}_{\mu\nu}$ do not match. The first term arises from the propagator of non-Lorentz violating modes, while all the other terms correspond to contributions from propagator of Lorentz violating modes. Hence, the quantum equivalence holds along non-Lorentz violating modes but not along Lorentz violating modes involving $b_{\mu\nu}$. This conclusion is validated by the results of Ref. \cite{seifert2010a}, where it was shown that when there are topologically nontrivial monopole-like solutions of the spontaneous symmetry breaking equations, the interaction with gravity of the vector and tensor theories are different.

\section{Conclusion}
Study of spontaneous Lorentz violation with rank-2 antisymmetric tensor is interesting because of the possibility of rich phenomenological signals of SME in future experiments. Since antisymmetric tensor fields are likely to play s significant role in the early universe cosmology, studying their quantum aspect is a natural extension of classical analyses. In a past study \cite{aashish2018b}, it was found that issues like quantum equivalence are difficult to address in a general curved spacetime. In the present work, this problem is overcome by adopting a perturbative approach to evaluating effective action, that is also general enough to be applied to more complicated models including interaction terms.

We quantized a simple action of an antisymmetric tensor field with a nonzero vev driving potential term that introduces spontaneous Lorentz violation, using a covariant effective action approach at one-loop. The one-loop corrections were calculated in a nearly flat spacetime, at $O(\kappa\hbar)$. We revisited the issue of quantum equivalence, and found that for the non-Lorentz-violating modes (independent of vev $b_{\mu\nu}$), antisymmetric tensor field is quantum-equivalent to a vector field. However, contributions from the Lorentz violating part of the propagator leads to different terms in effective actions, and as a result, $\Delta\Gamma = \Gamma_1 - \tilde{\Gamma}_1 \neq 0$, i.e. the theories are not quantum equivalent.


\begin{acknowledgments}
 This work was partially funded by DST (Govt. of India), Grant No. SERB/PHY/2017041. 
\end{acknowledgments}

\appendix*

\section{Projection operators for $B_{\mu\nu}$}
The basic projection operators for an antisymmetric tensor are defined as \cite{colatto2004},
\begin{equation} \begin{array}{rcl} \hspace{-0.8pc}P^{(1)}_{\mu\nu,\alpha\beta}&=&\displaystyle\frac{1}{2}(\theta_{\mu\alpha}\theta_{\nu\beta}-\theta_{\mu\beta}\theta_{\nu\alpha}),\\ \hspace{-0.8pc}P^{(2)}_{\mu\nu,\alpha\beta}&=&\displaystyle\frac{1}{4}(\theta_{\mu\alpha}\omega_{\nu\beta}-\theta_{\nu\alpha}\omega_{\mu\beta}-\theta_{\mu\beta}\omega_{\nu\alpha}+\theta_{\nu\beta}\omega_{\mu\alpha}), \end{array} \end{equation}
where,
\begin{equation} \theta_{\mu\nu}=\eta_{\mu\nu}-\omega_{\mu\nu},\quad \omega_{\mu\nu}=\frac{\partial_{\mu}\partial_{\nu}}{\Box}, \end{equation}
are the longitudinal and transverse projection operators along the momentum. To account for the Lorentz violation induced by nonzero vev, four new operators need to be introduced as follows \cite{maluf2019}:
\begin{eqnarray} 
P^{(3)}_{\mu\nu,\alpha\beta}&=& P^{\perp}_{\mu\nu, \alpha\beta},\\
P^{(4)}_{\mu\nu,\alpha\beta}&=& \frac{1}{2}\left( \omega_{\mu\lambda} \,P^{\parallel}_{\nu\lambda, \alpha\beta} - \omega_{\nu\lambda}\, P^{\parallel}_{\mu\lambda, \alpha\beta} \right),\\
P^{(5)}_{\mu\nu,\alpha\beta}&=& \frac{1}{2}\left( \omega_{\alpha\lambda} \,P^{\parallel}_{\mu\nu, \beta\lambda} - \omega_{\beta\lambda}\, P^{\parallel}_{\mu\nu, \alpha\lambda} \right),\\
P^{(6)}_{\mu\nu,\alpha\beta}&=& \frac{1}{4}\left( \omega_{\mu\alpha} \,P^{\parallel}_{\nu\rho, \beta\sigma}\,\omega^{\rho\sigma} - \omega_{\nu\alpha}\, P^{\parallel}_{\mu\rho, \beta\sigma}\,\omega^{\rho\sigma}\right. \nonumber\\ && {}-\,\left.\omega_{\mu\beta}\, P^{\parallel}_{\nu\rho, \alpha\sigma}\,\omega^{\rho\sigma} + \omega_{\nu\beta}\, P^{\parallel}_{\mu\rho, \alpha\sigma}\,\omega^{\rho\sigma} \right).
 \end{eqnarray}
The operators $P^{(1)}_{\mu\nu,\alpha\beta}, \cdots , P^{(6)}_{\mu\nu,\alpha\beta}$ obey a closed algebra \cite{maluf2019}. 

The identity element is given by,
\begin{equation} 
\mathcal{I}_{\mu\nu, \alpha\beta} = \frac{1}{2}(\eta_{\mu\alpha}\eta_{\nu\beta}-\eta_{\mu\beta}\eta_{\nu\alpha})=\left[P^{(1)}+P^{(2)}\right]_{\mu\nu,\alpha\beta}. 
\end{equation}


\bibliography{ref,references}

\begin{thebibliography}{53}%
\makeatletter
\providecommand \@ifxundefined [1]{%
 \@ifx{#1\undefined}
}%
\providecommand \@ifnum [1]{%
 \ifnum #1\expandafter \@firstoftwo
 \else \expandafter \@secondoftwo
 \fi
}%
\providecommand \@ifx [1]{%
 \ifx #1\expandafter \@firstoftwo
 \else \expandafter \@secondoftwo
 \fi
}%
\providecommand \natexlab [1]{#1}%
\providecommand \enquote  [1]{``#1''}%
\providecommand \bibnamefont  [1]{#1}%
\providecommand \bibfnamefont [1]{#1}%
\providecommand \citenamefont [1]{#1}%
\providecommand \href@noop [0]{\@secondoftwo}%
\providecommand \href [0]{\begingroup \@sanitize@url \@href}%
\providecommand \@href[1]{\@@startlink{#1}\@@href}%
\providecommand \@@href[1]{\endgroup#1\@@endlink}%
\providecommand \@sanitize@url [0]{\catcode `\\12\catcode `\$12\catcode
  `\&12\catcode `\#12\catcode `\^12\catcode `\_12\catcode `\%12\relax}%
\providecommand \@@startlink[1]{}%
\providecommand \@@endlink[0]{}%
\providecommand \url  [0]{\begingroup\@sanitize@url \@url }%
\providecommand \@url [1]{\endgroup\@href {#1}{\urlprefix }}%
\providecommand \urlprefix  [0]{URL }%
\providecommand \Eprint [0]{\href }%
\providecommand \doibase [0]{http://dx.doi.org/}%
\providecommand \selectlanguage [0]{\@gobble}%
\providecommand \bibinfo  [0]{\@secondoftwo}%
\providecommand \bibfield  [0]{\@secondoftwo}%
\providecommand \translation [1]{[#1]}%
\providecommand \BibitemOpen [0]{}%
\providecommand \bibitemStop [0]{}%
\providecommand \bibitemNoStop [0]{.\EOS\space}%
\providecommand \EOS [0]{\spacefactor3000\relax}%
\providecommand \BibitemShut  [1]{\csname bibitem#1\endcsname}%
\let\auto@bib@innerbib\@empty
\bibitem [{\citenamefont {Zimmermann}(2018)}]{zimmermann2018}%
  \BibitemOpen
  \bibfield  {author} {\bibinfo {author} {\bibfnamefont {F.}~\bibnamefont
  {Zimmermann}},\ }\href {\doibase https://doi.org/10.1016/j.nima.2018.01.034}
  {\bibfield  {journal} {\bibinfo  {journal} {Nuclear Instruments and Methods
  in Physics Research Section A: Accelerators, Spectrometers, Detectors and
  Associated Equipment}\ }\textbf {\bibinfo {volume} {909}},\ \bibinfo {pages}
  {33 } (\bibinfo {year} {2018})},\ \bibinfo {note} {3rd European Advanced
  Accelerator Concepts workshop (EAAC2017)}\BibitemShut {NoStop}%
\bibitem [{\citenamefont {Kosteleck\'y}\ and\ \citenamefont
  {Potting}(1995)}]{kostelecky1995}%
  \BibitemOpen
  \bibfield  {author} {\bibinfo {author} {\bibfnamefont {V.~A.}\ \bibnamefont
  {Kosteleck\'y}}\ and\ \bibinfo {author} {\bibfnamefont {R.}~\bibnamefont
  {Potting}},\ }\href {\doibase 10.1103/PhysRevD.51.3923} {\bibfield  {journal}
  {\bibinfo  {journal} {Phys. Rev. D}\ }\textbf {\bibinfo {volume} {51}},\
  \bibinfo {pages} {3923} (\bibinfo {year} {1995})}\BibitemShut {NoStop}%
\bibitem [{\citenamefont {Colladay}\ and\ \citenamefont
  {Kosteleck\'y}(1997)}]{colladay1997}%
  \BibitemOpen
  \bibfield  {author} {\bibinfo {author} {\bibfnamefont {D.}~\bibnamefont
  {Colladay}}\ and\ \bibinfo {author} {\bibfnamefont {V.~A.}\ \bibnamefont
  {Kosteleck\'y}},\ }\href {\doibase 10.1103/PhysRevD.55.6760} {\bibfield
  {journal} {\bibinfo  {journal} {Phys. Rev. D}\ }\textbf {\bibinfo {volume}
  {55}},\ \bibinfo {pages} {6760} (\bibinfo {year} {1997})}\BibitemShut
  {NoStop}%
\bibitem [{\citenamefont {Colladay}\ and\ \citenamefont
  {Kosteleck\'y}(1998)}]{colladay1998}%
  \BibitemOpen
  \bibfield  {author} {\bibinfo {author} {\bibfnamefont {D.}~\bibnamefont
  {Colladay}}\ and\ \bibinfo {author} {\bibfnamefont {V.~A.}\ \bibnamefont
  {Kosteleck\'y}},\ }\href {\doibase 10.1103/PhysRevD.58.116002} {\bibfield
  {journal} {\bibinfo  {journal} {Phys. Rev. D}\ }\textbf {\bibinfo {volume}
  {58}},\ \bibinfo {pages} {116002} (\bibinfo {year} {1998})}\BibitemShut
  {NoStop}%
\bibitem [{\citenamefont {Kosteleck\'y}(2004)}]{kostelecky2004}%
  \BibitemOpen
  \bibfield  {author} {\bibinfo {author} {\bibfnamefont {V.~A.}\ \bibnamefont
  {Kosteleck\'y}},\ }\href {\doibase 10.1103/PhysRevD.69.105009} {\bibfield
  {journal} {\bibinfo  {journal} {Phys. Rev. D}\ }\textbf {\bibinfo {volume}
  {69}},\ \bibinfo {pages} {105009} (\bibinfo {year} {2004})}\BibitemShut
  {NoStop}%
\bibitem [{\citenamefont {Bluhm}(2006)}]{bluhm2006}%
  \BibitemOpen
  \bibfield  {author} {\bibinfo {author} {\bibfnamefont {R.}~\bibnamefont
  {Bluhm}},\ }\enquote {\bibinfo {title} {Overview of the standard model
  extension: Implications and phenomenology of lorentz violation},}\ in\ \href
  {\doibase 10.1007/3-540-34523-X_8} {\emph {\bibinfo {booktitle} {Special
  Relativity: Will it Survive the Next 101 Years?}}},\ \bibinfo {editor}
  {edited by\ \bibinfo {editor} {\bibfnamefont {J.}~\bibnamefont {Ehlers}}\
  and\ \bibinfo {editor} {\bibfnamefont {C.}~\bibnamefont {L{\"a}mmerzahl}}}\
  (\bibinfo  {publisher} {Springer Berlin Heidelberg},\ \bibinfo {address}
  {Berlin, Heidelberg},\ \bibinfo {year} {2006})\ pp.\ \bibinfo {pages}
  {191--226}\BibitemShut {NoStop}%
\bibitem [{\citenamefont {Kosteleck\'y}\ and\ \citenamefont
  {Samuel}(1989)}]{kostelecky1989b}%
  \BibitemOpen
  \bibfield  {author} {\bibinfo {author} {\bibfnamefont {V.~A.}\ \bibnamefont
  {Kosteleck\'y}}\ and\ \bibinfo {author} {\bibfnamefont {S.}~\bibnamefont
  {Samuel}},\ }\href {\doibase 10.1103/PhysRevD.39.683} {\bibfield  {journal}
  {\bibinfo  {journal} {Phys. Rev. D}\ }\textbf {\bibinfo {volume} {39}},\
  \bibinfo {pages} {683} (\bibinfo {year} {1989})}\BibitemShut {NoStop}%
\bibitem [{\citenamefont {Kostelecký}\ and\ \citenamefont
  {Potting}(1991)}]{kostelecky1991a}%
  \BibitemOpen
  \bibfield  {author} {\bibinfo {author} {\bibfnamefont {V.~A.}\ \bibnamefont
  {Kostelecký}}\ and\ \bibinfo {author} {\bibfnamefont {R.}~\bibnamefont
  {Potting}},\ }\href {\doibase https://doi.org/10.1016/0550-3213(91)90071-5}
  {\bibfield  {journal} {\bibinfo  {journal} {Nuclear Physics B}\ }\textbf
  {\bibinfo {volume} {359}},\ \bibinfo {pages} {545 } (\bibinfo {year}
  {1991})}\BibitemShut {NoStop}%
\bibitem [{\citenamefont {Kosteleck\'y}\ and\ \citenamefont
  {Samuel}(1991)}]{kostelecky1991b}%
  \BibitemOpen
  \bibfield  {author} {\bibinfo {author} {\bibfnamefont {V.~A.}\ \bibnamefont
  {Kosteleck\'y}}\ and\ \bibinfo {author} {\bibfnamefont {S.}~\bibnamefont
  {Samuel}},\ }\href {\doibase 10.1103/PhysRevLett.66.1811} {\bibfield
  {journal} {\bibinfo  {journal} {Phys. Rev. Lett.}\ }\textbf {\bibinfo
  {volume} {66}},\ \bibinfo {pages} {1811} (\bibinfo {year}
  {1991})}\BibitemShut {NoStop}%
\bibitem [{\citenamefont {Kostelecký}\ and\ \citenamefont
  {Potting}(1996)}]{kostelecky1996}%
  \BibitemOpen
  \bibfield  {author} {\bibinfo {author} {\bibfnamefont {V.~A.}\ \bibnamefont
  {Kostelecký}}\ and\ \bibinfo {author} {\bibfnamefont {R.}~\bibnamefont
  {Potting}},\ }\href {\doibase https://doi.org/10.1016/0370-2693(96)00589-8}
  {\bibfield  {journal} {\bibinfo  {journal} {Physics Letters B}\ }\textbf
  {\bibinfo {volume} {381}},\ \bibinfo {pages} {89 } (\bibinfo {year}
  {1996})}\BibitemShut {NoStop}%
\bibitem [{\citenamefont {Kosteleck\'y}\ and\ \citenamefont
  {Potting}(2001)}]{kostelecky2001}%
  \BibitemOpen
  \bibfield  {author} {\bibinfo {author} {\bibfnamefont {V.~A.}\ \bibnamefont
  {Kosteleck\'y}}\ and\ \bibinfo {author} {\bibfnamefont {R.}~\bibnamefont
  {Potting}},\ }\href {\doibase 10.1103/PhysRevD.63.046007} {\bibfield
  {journal} {\bibinfo  {journal} {Phys. Rev. D}\ }\textbf {\bibinfo {volume}
  {63}},\ \bibinfo {pages} {046007} (\bibinfo {year} {2001})}\BibitemShut
  {NoStop}%
\bibitem [{\citenamefont {Gambini}\ and\ \citenamefont
  {Pullin}(1999)}]{gambini1999}%
  \BibitemOpen
  \bibfield  {author} {\bibinfo {author} {\bibfnamefont {R.}~\bibnamefont
  {Gambini}}\ and\ \bibinfo {author} {\bibfnamefont {J.}~\bibnamefont
  {Pullin}},\ }\href {\doibase 10.1103/PhysRevD.59.124021} {\bibfield
  {journal} {\bibinfo  {journal} {Phys. Rev. D}\ }\textbf {\bibinfo {volume}
  {59}},\ \bibinfo {pages} {124021} (\bibinfo {year} {1999})}\BibitemShut
  {NoStop}%
\bibitem [{\citenamefont {Alfaro}\ \emph {et~al.}(2002)\citenamefont {Alfaro},
  \citenamefont {Morales-T\'ecotl},\ and\ \citenamefont
  {Urrutia}}]{alfaro2002}%
  \BibitemOpen
  \bibfield  {author} {\bibinfo {author} {\bibfnamefont {J.}~\bibnamefont
  {Alfaro}}, \bibinfo {author} {\bibfnamefont {H.~A.}\ \bibnamefont
  {Morales-T\'ecotl}}, \ and\ \bibinfo {author} {\bibfnamefont {L.~F.}\
  \bibnamefont {Urrutia}},\ }\href {\doibase 10.1103/PhysRevD.66.124006}
  {\bibfield  {journal} {\bibinfo  {journal} {Phys. Rev. D}\ }\textbf {\bibinfo
  {volume} {66}},\ \bibinfo {pages} {124006} (\bibinfo {year}
  {2002})}\BibitemShut {NoStop}%
\bibitem [{\citenamefont {Sudarsky}\ \emph {et~al.}(2002)\citenamefont
  {Sudarsky}, \citenamefont {Urrutia},\ and\ \citenamefont
  {Vucetich}}]{sudarsky2002}%
  \BibitemOpen
  \bibfield  {author} {\bibinfo {author} {\bibfnamefont {D.}~\bibnamefont
  {Sudarsky}}, \bibinfo {author} {\bibfnamefont {L.}~\bibnamefont {Urrutia}}, \
  and\ \bibinfo {author} {\bibfnamefont {H.}~\bibnamefont {Vucetich}},\ }\href
  {\doibase 10.1103/PhysRevLett.89.231301} {\bibfield  {journal} {\bibinfo
  {journal} {Phys. Rev. Lett.}\ }\textbf {\bibinfo {volume} {89}},\ \bibinfo
  {pages} {231301} (\bibinfo {year} {2002})}\BibitemShut {NoStop}%
\bibitem [{\citenamefont {Sudarsky}\ \emph {et~al.}(2003)\citenamefont
  {Sudarsky}, \citenamefont {Urrutia},\ and\ \citenamefont
  {Vucetich}}]{sudarsky2003}%
  \BibitemOpen
  \bibfield  {author} {\bibinfo {author} {\bibfnamefont {D.}~\bibnamefont
  {Sudarsky}}, \bibinfo {author} {\bibfnamefont {L.}~\bibnamefont {Urrutia}}, \
  and\ \bibinfo {author} {\bibfnamefont {H.}~\bibnamefont {Vucetich}},\ }\href
  {\doibase 10.1103/PhysRevD.68.024010} {\bibfield  {journal} {\bibinfo
  {journal} {Phys. Rev. D}\ }\textbf {\bibinfo {volume} {68}},\ \bibinfo
  {pages} {024010} (\bibinfo {year} {2003})}\BibitemShut {NoStop}%
\bibitem [{\citenamefont {Myers}\ and\ \citenamefont
  {Pospelov}(2003)}]{myers2003}%
  \BibitemOpen
  \bibfield  {author} {\bibinfo {author} {\bibfnamefont {R.~C.}\ \bibnamefont
  {Myers}}\ and\ \bibinfo {author} {\bibfnamefont {M.}~\bibnamefont
  {Pospelov}},\ }\href {\doibase 10.1103/PhysRevLett.90.211601} {\bibfield
  {journal} {\bibinfo  {journal} {Phys. Rev. Lett.}\ }\textbf {\bibinfo
  {volume} {90}},\ \bibinfo {pages} {211601} (\bibinfo {year}
  {2003})}\BibitemShut {NoStop}%
\bibitem [{\citenamefont {Rohm}\ and\ \citenamefont {Witten}(1986)}]{rohm1986}%
  \BibitemOpen
  \bibfield  {author} {\bibinfo {author} {\bibfnamefont {R.}~\bibnamefont
  {Rohm}}\ and\ \bibinfo {author} {\bibfnamefont {E.}~\bibnamefont {Witten}},\
  }\href {\doibase https://doi.org/10.1016/0003-4916(86)90099-0} {\bibfield
  {journal} {\bibinfo  {journal} {Annals of Physics}\ }\textbf {\bibinfo
  {volume} {170}},\ \bibinfo {pages} {454 } (\bibinfo {year}
  {1986})}\BibitemShut {NoStop}%
\bibitem [{\citenamefont {Ghezelbash}(2009)}]{ghezelbash2009}%
  \BibitemOpen
  \bibfield  {author} {\bibinfo {author} {\bibfnamefont {A.~M.}\ \bibnamefont
  {Ghezelbash}},\ }\href {\doibase 10.1088/1126-6708/2009/08/045} {\bibfield
  {journal} {\bibinfo  {journal} {JHEP}\ }\textbf {\bibinfo {volume} {08}},\
  \bibinfo {pages} {045} (\bibinfo {year} {2009})},\ \Eprint
  {http://arxiv.org/abs/0901.1670} {arXiv:0901.1670 [hep-th]} \BibitemShut
  {NoStop}%
\bibitem [{\citenamefont {Koivisto}\ \emph {et~al.}(2009)\citenamefont
  {Koivisto}, \citenamefont {Mota},\ and\ \citenamefont
  {Pitrou}}]{koivisto2009a}%
  \BibitemOpen
  \bibfield  {author} {\bibinfo {author} {\bibfnamefont {T.~S.}\ \bibnamefont
  {Koivisto}}, \bibinfo {author} {\bibfnamefont {D.~F.}\ \bibnamefont {Mota}},
  \ and\ \bibinfo {author} {\bibfnamefont {C.}~\bibnamefont {Pitrou}},\ }\href
  {http://stacks.iop.org/1126-6708/2009/i=09/a=092} {\bibfield  {journal}
  {\bibinfo  {journal} {Journal of High Energy Physics}\ }\textbf {\bibinfo
  {volume} {2009}},\ \bibinfo {pages} {092} (\bibinfo {year}
  {2009})}\BibitemShut {NoStop}%
\bibitem [{\citenamefont {Aashish}\ \emph {et~al.}(2018)\citenamefont
  {Aashish}, \citenamefont {Padhy}, \citenamefont {Panda},\ and\ \citenamefont
  {Rana}}]{aashish2018c}%
  \BibitemOpen
  \bibfield  {author} {\bibinfo {author} {\bibfnamefont {S.}~\bibnamefont
  {Aashish}}, \bibinfo {author} {\bibfnamefont {A.}~\bibnamefont {Padhy}},
  \bibinfo {author} {\bibfnamefont {S.}~\bibnamefont {Panda}}, \ and\ \bibinfo
  {author} {\bibfnamefont {A.}~\bibnamefont {Rana}},\ }\href@noop {} {\bibfield
   {journal} {\bibinfo  {journal} {The European Physical Journal C}\ }\textbf
  {\bibinfo {volume} {78}},\ \bibinfo {pages} {887} (\bibinfo {year}
  {2018})}\BibitemShut {NoStop}%
\bibitem [{\citenamefont {Janssen}\ and\ \citenamefont
  {Prokopec}(2006)}]{prokopec2006}%
  \BibitemOpen
  \bibfield  {author} {\bibinfo {author} {\bibfnamefont {T.}~\bibnamefont
  {Janssen}}\ and\ \bibinfo {author} {\bibfnamefont {T.}~\bibnamefont
  {Prokopec}},\ }\href {http://stacks.iop.org/0264-9381/23/i=15/a=015}
  {\bibfield  {journal} {\bibinfo  {journal} {Classical and Quantum Gravity}\
  }\textbf {\bibinfo {volume} {23}},\ \bibinfo {pages} {4967} (\bibinfo {year}
  {2006})}\BibitemShut {NoStop}%
\bibitem [{\citenamefont {Aashish}\ \emph {et~al.}(2019)\citenamefont
  {Aashish}, \citenamefont {Padhy},\ and\ \citenamefont
  {Panda}}]{aashish2019a}%
  \BibitemOpen
  \bibfield  {author} {\bibinfo {author} {\bibfnamefont {S.}~\bibnamefont
  {Aashish}}, \bibinfo {author} {\bibfnamefont {A.}~\bibnamefont {Padhy}}, \
  and\ \bibinfo {author} {\bibfnamefont {S.}~\bibnamefont {Panda}},\
  }\href@noop {} {\  (\bibinfo {year} {2019})},\ \Eprint
  {http://arxiv.org/abs/1901.10959} {arXiv:1901.10959 [gr-qc]} \BibitemShut
  {NoStop}%
\bibitem [{\citenamefont {Elizalde}\ \emph {et~al.}(2019)\citenamefont
  {Elizalde}, \citenamefont {Odintsov}, \citenamefont {Paul},\ and\
  \citenamefont {G\'omez}}]{elizalde2018}%
  \BibitemOpen
  \bibfield  {author} {\bibinfo {author} {\bibfnamefont {E.}~\bibnamefont
  {Elizalde}}, \bibinfo {author} {\bibfnamefont {S.~D.}\ \bibnamefont
  {Odintsov}}, \bibinfo {author} {\bibfnamefont {T.}~\bibnamefont {Paul}}, \
  and\ \bibinfo {author} {\bibfnamefont {D.~S.-C.}\ \bibnamefont {G\'omez}},\
  }\href {\doibase 10.1103/PhysRevD.99.063506} {\bibfield  {journal} {\bibinfo
  {journal} {Phys. Rev. D}\ }\textbf {\bibinfo {volume} {99}},\ \bibinfo
  {pages} {063506} (\bibinfo {year} {2019})}\BibitemShut {NoStop}%
\bibitem [{\citenamefont {Assunção}\ \emph {et~al.}(2019)\citenamefont
  {Assunção}, \citenamefont {Mariz}, \citenamefont {Nascimento},\ and\
  \citenamefont {Petrov}}]{petrov2019}%
  \BibitemOpen
  \bibfield  {author} {\bibinfo {author} {\bibfnamefont {J.~F.}\ \bibnamefont
  {Assunção}}, \bibinfo {author} {\bibfnamefont {T.}~\bibnamefont {Mariz}},
  \bibinfo {author} {\bibfnamefont {J.~R.}\ \bibnamefont {Nascimento}}, \ and\
  \bibinfo {author} {\bibfnamefont {A.~{\relax Yu}.}\ \bibnamefont {Petrov}},\
  }\href@noop {} {\  (\bibinfo {year} {2019})},\ \Eprint
  {http://arxiv.org/abs/1902.10592} {arXiv:1902.10592 [hep-th]} \BibitemShut
  {NoStop}%
\bibitem [{\citenamefont {Altschul}\ \emph {et~al.}(2010)\citenamefont
  {Altschul}, \citenamefont {Bailey},\ and\ \citenamefont
  {Kostelecky}}]{altschul2010}%
  \BibitemOpen
  \bibfield  {author} {\bibinfo {author} {\bibfnamefont {B.}~\bibnamefont
  {Altschul}}, \bibinfo {author} {\bibfnamefont {Q.~G.}\ \bibnamefont
  {Bailey}}, \ and\ \bibinfo {author} {\bibfnamefont {V.~A.}\ \bibnamefont
  {Kostelecky}},\ }\href {\doibase 10.1103/PhysRevD.81.065028} {\bibfield
  {journal} {\bibinfo  {journal} {Phys. Rev.}\ }\textbf {\bibinfo {volume}
  {D81}},\ \bibinfo {pages} {065028} (\bibinfo {year} {2010})},\ \Eprint
  {http://arxiv.org/abs/0912.4852} {arXiv:0912.4852 [gr-qc]} \BibitemShut
  {NoStop}%
\bibitem [{\citenamefont {DeWitt}(1964)}]{dewitt1964}%
  \BibitemOpen
  \bibfield  {author} {\bibinfo {author} {\bibfnamefont {B.~S.}\ \bibnamefont
  {DeWitt}},\ }\bibfield  {booktitle} {\emph {\bibinfo {booktitle}
  {{Relativité, Groupes et Topologie: Proceedings, Ecole d'été de Physique
  Théorique, Session XIII, Les Houches, France, Jul 1 - Aug 24, 1963}}},\
  }\href@noop {} {\bibfield  {journal} {\bibinfo  {journal} {Conf. Proc.}\
  }\textbf {\bibinfo {volume} {C630701}},\ \bibinfo {pages} {585} (\bibinfo
  {year} {1964})},\ \bibinfo {note} {[Les Houches Lect.
  Notes13,585(1964)]}\BibitemShut {NoStop}%
\bibitem [{\citenamefont {DeWitt}(1967{\natexlab{a}})}]{dewitt1967a}%
  \BibitemOpen
  \bibfield  {author} {\bibinfo {author} {\bibfnamefont {B.~S.}\ \bibnamefont
  {DeWitt}},\ }\href {\doibase 10.1103/PhysRev.160.1113} {\bibfield  {journal}
  {\bibinfo  {journal} {Phys. Rev.}\ }\textbf {\bibinfo {volume} {160}},\
  \bibinfo {pages} {1113} (\bibinfo {year} {1967}{\natexlab{a}})}\BibitemShut
  {NoStop}%
\bibitem [{\citenamefont {DeWitt}(1967{\natexlab{b}})}]{dewitt1967b}%
  \BibitemOpen
  \bibfield  {author} {\bibinfo {author} {\bibfnamefont {B.~S.}\ \bibnamefont
  {DeWitt}},\ }\href {\doibase 10.1103/PhysRev.162.1195} {\bibfield  {journal}
  {\bibinfo  {journal} {Phys. Rev.}\ }\textbf {\bibinfo {volume} {162}},\
  \bibinfo {pages} {1195} (\bibinfo {year} {1967}{\natexlab{b}})}\BibitemShut
  {NoStop}%
\bibitem [{\citenamefont {DeWitt}(1967{\natexlab{c}})}]{dewitt1967c}%
  \BibitemOpen
  \bibfield  {author} {\bibinfo {author} {\bibfnamefont {B.~S.}\ \bibnamefont
  {DeWitt}},\ }\href {\doibase 10.1103/PhysRev.162.1239} {\bibfield  {journal}
  {\bibinfo  {journal} {Phys. Rev.}\ }\textbf {\bibinfo {volume} {162}},\
  \bibinfo {pages} {1239} (\bibinfo {year} {1967}{\natexlab{c}})}\BibitemShut
  {NoStop}%
\bibitem [{\citenamefont {Vilkovisky}(1984{\natexlab{a}})}]{vilkovisky1984a}%
  \BibitemOpen
  \bibfield  {author} {\bibinfo {author} {\bibfnamefont {G.~A.}\ \bibnamefont
  {Vilkovisky}},\ }\href {\doibase 10.1016/0550-3213(84)90228-1} {\bibfield
  {journal} {\bibinfo  {journal} {Nucl. Phys.}\ }\textbf {\bibinfo {volume}
  {B234}},\ \bibinfo {pages} {125} (\bibinfo {year}
  {1984}{\natexlab{a}})}\BibitemShut {NoStop}%
\bibitem [{\citenamefont {Vilkovisky}(1984{\natexlab{b}})}]{vilkovisky1984b}%
  \BibitemOpen
  \bibfield  {author} {\bibinfo {author} {\bibfnamefont {G.~A.}\ \bibnamefont
  {Vilkovisky}},\ }\href@noop {} {\bibfield  {journal} {\bibinfo  {journal} {In
  Christensen, S.M. ( Ed.): Quantum Theory Of Gravity}\ ,\ \bibinfo {pages}
  {169}} (\bibinfo {year} {1984}{\natexlab{b}})}\BibitemShut {NoStop}%
\bibitem [{\citenamefont {Parker}\ and\ \citenamefont
  {Toms}(2009)}]{parker2009}%
  \BibitemOpen
  \bibfield  {author} {\bibinfo {author} {\bibfnamefont {L.~E.}\ \bibnamefont
  {Parker}}\ and\ \bibinfo {author} {\bibfnamefont {D.}~\bibnamefont {Toms}},\
  }\href {\doibase 10.1017/CBO9780511813924} {\emph {\bibinfo {title} {{Quantum
  Field Theory in Curved Spacetime}}}},\ Cambridge Monographs on Mathematical
  Physics\ (\bibinfo  {publisher} {Cambridge University Press},\ \bibinfo
  {year} {2009})\BibitemShut {NoStop}%
\bibitem [{\citenamefont {Buchbinder}\ \emph {et~al.}(2008)\citenamefont
  {Buchbinder}, \citenamefont {Kirillova},\ and\ \citenamefont
  {Pletnev}}]{buchbinder2008}%
  \BibitemOpen
  \bibfield  {author} {\bibinfo {author} {\bibfnamefont {I.~L.}\ \bibnamefont
  {Buchbinder}}, \bibinfo {author} {\bibfnamefont {E.~N.}\ \bibnamefont
  {Kirillova}}, \ and\ \bibinfo {author} {\bibfnamefont {N.~G.}\ \bibnamefont
  {Pletnev}},\ }\href {\doibase 10.1103/PhysRevD.78.084024} {\bibfield
  {journal} {\bibinfo  {journal} {Phys. Rev.}\ }\textbf {\bibinfo {volume}
  {D78}},\ \bibinfo {pages} {084024} (\bibinfo {year} {2008})},\ \Eprint
  {http://arxiv.org/abs/0806.3505} {arXiv:0806.3505 [hep-th]} \BibitemShut
  {NoStop}%
\bibitem [{\citenamefont {Aashish}\ and\ \citenamefont
  {Panda}(2018{\natexlab{a}})}]{aashish2018b}%
  \BibitemOpen
  \bibfield  {author} {\bibinfo {author} {\bibfnamefont {S.}~\bibnamefont
  {Aashish}}\ and\ \bibinfo {author} {\bibfnamefont {S.}~\bibnamefont
  {Panda}},\ }\href@noop {} {\  (\bibinfo {year} {2018}{\natexlab{a}})},\
  \Eprint {http://arxiv.org/abs/1806.08194} {arXiv:1806.08194 [gr-qc]}
  \BibitemShut {NoStop}%
\bibitem [{\citenamefont {Seifert}(2010{\natexlab{a}})}]{seifert2010a}%
  \BibitemOpen
  \bibfield  {author} {\bibinfo {author} {\bibfnamefont {M.~D.}\ \bibnamefont
  {Seifert}},\ }\href {\doibase 10.1103/PhysRevLett.105.201601} {\bibfield
  {journal} {\bibinfo  {journal} {Phys. Rev. Lett.}\ }\textbf {\bibinfo
  {volume} {105}},\ \bibinfo {pages} {201601} (\bibinfo {year}
  {2010}{\natexlab{a}})}\BibitemShut {NoStop}%
\bibitem [{\citenamefont {Bluhm}\ and\ \citenamefont
  {Kosteleck\'y}(2005)}]{bluhm2005}%
  \BibitemOpen
  \bibfield  {author} {\bibinfo {author} {\bibfnamefont {R.}~\bibnamefont
  {Bluhm}}\ and\ \bibinfo {author} {\bibfnamefont {V.~A.}\ \bibnamefont
  {Kosteleck\'y}},\ }\href {\doibase 10.1103/PhysRevD.71.065008} {\bibfield
  {journal} {\bibinfo  {journal} {Phys. Rev. D}\ }\textbf {\bibinfo {volume}
  {71}},\ \bibinfo {pages} {065008} (\bibinfo {year} {2005})}\BibitemShut
  {NoStop}%
\bibitem [{\citenamefont {Seifert}(2010{\natexlab{b}})}]{seifert2010b}%
  \BibitemOpen
  \bibfield  {author} {\bibinfo {author} {\bibfnamefont {M.~D.}\ \bibnamefont
  {Seifert}},\ }\href {\doibase 10.1103/PhysRevD.82.125015} {\bibfield
  {journal} {\bibinfo  {journal} {Phys. Rev. D}\ }\textbf {\bibinfo {volume}
  {82}},\ \bibinfo {pages} {125015} (\bibinfo {year}
  {2010}{\natexlab{b}})}\BibitemShut {NoStop}%
\bibitem [{\citenamefont {Buchbinder}\ \emph {et~al.}(2007)\citenamefont
  {Buchbinder}, \citenamefont {de~Berredo-Peixoto},\ and\ \citenamefont
  {Shapiro}}]{buchbinder2007}%
  \BibitemOpen
  \bibfield  {author} {\bibinfo {author} {\bibfnamefont {I.~L.}\ \bibnamefont
  {Buchbinder}}, \bibinfo {author} {\bibfnamefont {G.}~\bibnamefont
  {de~Berredo-Peixoto}}, \ and\ \bibinfo {author} {\bibfnamefont {I.~L.}\
  \bibnamefont {Shapiro}},\ }\href {\doibase 10.1016/j.physletb.2007.04.039}
  {\bibfield  {journal} {\bibinfo  {journal} {Phys. Lett.}\ }\textbf {\bibinfo
  {volume} {B649}},\ \bibinfo {pages} {454} (\bibinfo {year} {2007})},\ \Eprint
  {http://arxiv.org/abs/hep-th/0703189} {arXiv:hep-th/0703189 [HEP-TH]}
  \BibitemShut {NoStop}%
\bibitem [{\citenamefont {St{\"u}ckelberg}(1957)}]{stuckelberg1957}%
  \BibitemOpen
  \bibfield  {author} {\bibinfo {author} {\bibfnamefont {E.}~\bibnamefont
  {St{\"u}ckelberg}},\ }\href@noop {} {\bibfield  {journal} {\bibinfo
  {journal} {Helv. Phys. Acta}\ }\textbf {\bibinfo {volume} {30}},\ \bibinfo
  {pages} {209} (\bibinfo {year} {1957})}\BibitemShut {NoStop}%
\bibitem [{\citenamefont {Buchbinder}\ \emph {et~al.}(1992)\citenamefont
  {Buchbinder}, \citenamefont {Odintsov},\ and\ \citenamefont
  {Shapiro}}]{buchbinder1992}%
  \BibitemOpen
  \bibfield  {author} {\bibinfo {author} {\bibfnamefont {I.~L.}\ \bibnamefont
  {Buchbinder}}, \bibinfo {author} {\bibfnamefont {S.~D.}\ \bibnamefont
  {Odintsov}}, \ and\ \bibinfo {author} {\bibfnamefont {I.~L.}\ \bibnamefont
  {Shapiro}},\ }\href@noop {} {\emph {\bibinfo {title} {{Effective action in
  quantum gravity}}}}\ (\bibinfo  {publisher} {Taylor \& Francis Group},\
  \bibinfo {year} {1992})\BibitemShut {NoStop}%
\bibitem [{\citenamefont {Aashish}\ and\ \citenamefont
  {Panda}(2018{\natexlab{b}})}]{aashish2018a}%
  \BibitemOpen
  \bibfield  {author} {\bibinfo {author} {\bibfnamefont {S.}~\bibnamefont
  {Aashish}}\ and\ \bibinfo {author} {\bibfnamefont {S.}~\bibnamefont
  {Panda}},\ }\href {\doibase 10.1103/PhysRevD.97.125005} {\bibfield  {journal}
  {\bibinfo  {journal} {Phys. Rev. D}\ }\textbf {\bibinfo {volume} {97}},\
  \bibinfo {pages} {125005} (\bibinfo {year} {2018}{\natexlab{b}})},\ \Eprint
  {http://arxiv.org/abs/1803.10157} {arXiv:1803.10157 [gr-qc]} \BibitemShut
  {NoStop}%
\bibitem [{\citenamefont {Buchbinder}\ and\ \citenamefont
  {Kuzenko}(1988)}]{buchbinder1988}%
  \BibitemOpen
  \bibfield  {author} {\bibinfo {author} {\bibfnamefont {I.}~\bibnamefont
  {Buchbinder}}\ and\ \bibinfo {author} {\bibfnamefont {S.}~\bibnamefont
  {Kuzenko}},\ }\href {\doibase https://doi.org/10.1016/0550-3213(88)90047-8}
  {\bibfield  {journal} {\bibinfo  {journal} {Nuclear Physics B}\ }\textbf
  {\bibinfo {volume} {308}},\ \bibinfo {pages} {162 } (\bibinfo {year}
  {1988})}\BibitemShut {NoStop}%
\bibitem [{\citenamefont {Brizuela}\ \emph {et~al.}(2009)\citenamefont
  {Brizuela}, \citenamefont {Mart{\'i}n-Garc{\'i}a},\ and\ \citenamefont
  {Mena~Marug{\'a}n}}]{xpert}%
  \BibitemOpen
  \bibfield  {author} {\bibinfo {author} {\bibfnamefont {D.}~\bibnamefont
  {Brizuela}}, \bibinfo {author} {\bibfnamefont {J.~M.}\ \bibnamefont
  {Mart{\'i}n-Garc{\'i}a}}, \ and\ \bibinfo {author} {\bibfnamefont {G.~A.}\
  \bibnamefont {Mena~Marug{\'a}n}},\ }\href {\doibase
  10.1007/s10714-009-0773-2} {\bibfield  {journal} {\bibinfo  {journal}
  {General Relativity and Gravitation}\ }\textbf {\bibinfo {volume} {41}},\
  \bibinfo {pages} {2415} (\bibinfo {year} {2009})}\BibitemShut {NoStop}%
\bibitem [{\citenamefont {Mart{\'i}n-Garc{\'i}a}()}]{xact}%
  \BibitemOpen
  \bibfield  {author} {\bibinfo {author} {\bibfnamefont {J.~M.}\ \bibnamefont
  {Mart{\'i}n-Garc{\'i}a}},\ }\href@noop {} {\enquote {\bibinfo {title} {xact:
  Efficient tensor computer algebra},}\ }\bibinfo {howpublished}
  {\url{http://www.xact.es}}\BibitemShut {NoStop}%
\bibitem [{\citenamefont {Barnes}(1965)}]{barnes1965}%
  \BibitemOpen
  \bibfield  {author} {\bibinfo {author} {\bibfnamefont {K.~J.}\ \bibnamefont
  {Barnes}},\ }\href {\doibase 10.1063/1.1704335} {\bibfield  {journal}
  {\bibinfo  {journal} {Journal of Mathematical Physics}\ }\textbf {\bibinfo
  {volume} {6}},\ \bibinfo {pages} {788} (\bibinfo {year} {1965})},\ \Eprint
  {http://arxiv.org/abs/https://doi.org/10.1063/1.1704335}
  {https://doi.org/10.1063/1.1704335} \BibitemShut {NoStop}%
\bibitem [{\citenamefont {Maluf}\ \emph {et~al.}(2019)\citenamefont {Maluf},
  \citenamefont {Filho}, \citenamefont {Cruz},\ and\ \citenamefont
  {Almeida}}]{maluf2019}%
  \BibitemOpen
  \bibfield  {author} {\bibinfo {author} {\bibfnamefont {R.~V.}\ \bibnamefont
  {Maluf}}, \bibinfo {author} {\bibfnamefont {A.~A.~A.}\ \bibnamefont {Filho}},
  \bibinfo {author} {\bibfnamefont {W.~T.}\ \bibnamefont {Cruz}}, \ and\
  \bibinfo {author} {\bibfnamefont {C.~A.~S.}\ \bibnamefont {Almeida}},\ }\href
  {\doibase 10.1209/0295-5075/124/61001} {\bibfield  {journal} {\bibinfo
  {journal} {{EPL} (Europhysics Letters)}\ }\textbf {\bibinfo {volume} {124}},\
  \bibinfo {pages} {61001} (\bibinfo {year} {2019})}\BibitemShut {NoStop}%
\bibitem [{\citenamefont {Bardin}\ and\ \citenamefont
  {Passarino}(1999)}]{bardin1999}%
  \BibitemOpen
  \bibfield  {author} {\bibinfo {author} {\bibfnamefont {D.~{\relax Yu}.}\
  \bibnamefont {Bardin}}\ and\ \bibinfo {author} {\bibfnamefont
  {G.}~\bibnamefont {Passarino}},\ }\href@noop {} {\emph {\bibinfo {title}
  {{The standard model in the making: Precision study of the electroweak
  interactions}}}}\ (\bibinfo {year} {1999})\BibitemShut {NoStop}%
\bibitem [{\citenamefont {Leibbrandt}(1975)}]{george1975}%
  \BibitemOpen
  \bibfield  {author} {\bibinfo {author} {\bibfnamefont {G.}~\bibnamefont
  {Leibbrandt}},\ }\href {\doibase 10.1103/RevModPhys.47.849} {\bibfield
  {journal} {\bibinfo  {journal} {Rev. Mod. Phys.}\ }\textbf {\bibinfo {volume}
  {47}},\ \bibinfo {pages} {849} (\bibinfo {year} {1975})}\BibitemShut
  {NoStop}%
\bibitem [{\citenamefont {Leibbrandt}(1987)}]{george1987}%
  \BibitemOpen
  \bibfield  {author} {\bibinfo {author} {\bibfnamefont {G.}~\bibnamefont
  {Leibbrandt}},\ }\href {\doibase 10.1103/RevModPhys.59.1067} {\bibfield
  {journal} {\bibinfo  {journal} {Rev. Mod. Phys.}\ }\textbf {\bibinfo {volume}
  {59}},\ \bibinfo {pages} {1067} (\bibinfo {year} {1987})}\BibitemShut
  {NoStop}%
\bibitem [{\citenamefont {Capper}\ and\ \citenamefont
  {Leibbrandt}(1982{\natexlab{a}})}]{capper1982a}%
  \BibitemOpen
  \bibfield  {author} {\bibinfo {author} {\bibfnamefont {D.~M.}\ \bibnamefont
  {Capper}}\ and\ \bibinfo {author} {\bibfnamefont {G.}~\bibnamefont
  {Leibbrandt}},\ }\href {\doibase 10.1103/PhysRevD.25.1002} {\bibfield
  {journal} {\bibinfo  {journal} {Phys. Rev. D}\ }\textbf {\bibinfo {volume}
  {25}},\ \bibinfo {pages} {1002} (\bibinfo {year}
  {1982}{\natexlab{a}})}\BibitemShut {NoStop}%
\bibitem [{\citenamefont {Capper}\ and\ \citenamefont
  {Leibbrandt}(1982{\natexlab{b}})}]{capper1982b}%
  \BibitemOpen
  \bibfield  {author} {\bibinfo {author} {\bibfnamefont {D.~M.}\ \bibnamefont
  {Capper}}\ and\ \bibinfo {author} {\bibfnamefont {G.}~\bibnamefont
  {Leibbrandt}},\ }\href {\doibase 10.1103/PhysRevD.25.1009} {\bibfield
  {journal} {\bibinfo  {journal} {Phys. Rev. D}\ }\textbf {\bibinfo {volume}
  {25}},\ \bibinfo {pages} {1009} (\bibinfo {year}
  {1982}{\natexlab{b}})}\BibitemShut {NoStop}%
\bibitem [{\citenamefont {Mackay}\ and\ \citenamefont
  {Toms}(2010)}]{mackay2010}%
  \BibitemOpen
  \bibfield  {author} {\bibinfo {author} {\bibfnamefont {P.~T.}\ \bibnamefont
  {Mackay}}\ and\ \bibinfo {author} {\bibfnamefont {D.~J.}\ \bibnamefont
  {Toms}},\ }\href {\doibase https://doi.org/10.1016/j.physletb.2009.12.032}
  {\bibfield  {journal} {\bibinfo  {journal} {Physics Letters B}\ }\textbf
  {\bibinfo {volume} {684}},\ \bibinfo {pages} {251 } (\bibinfo {year}
  {2010})}\BibitemShut {NoStop}%
\bibitem [{\citenamefont {Colatto}\ \emph {et~al.}(2004)\citenamefont
  {Colatto}, \citenamefont {Penna},\ and\ \citenamefont
  {Santos}}]{colatto2004}%
  \BibitemOpen
  \bibfield  {author} {\bibinfo {author} {\bibfnamefont {L.~P.}\ \bibnamefont
  {Colatto}}, \bibinfo {author} {\bibfnamefont {A.~L.~A.}\ \bibnamefont
  {Penna}}, \ and\ \bibinfo {author} {\bibfnamefont {W.~C.}\ \bibnamefont
  {Santos}},\ }\href {\doibase 10.1140/epjc/s2004-01865-6} {\bibfield
  {journal} {\bibinfo  {journal} {The European Physical Journal C - Particles
  and Fields}\ }\textbf {\bibinfo {volume} {36}},\ \bibinfo {pages} {79}
  (\bibinfo {year} {2004})}\BibitemShut {NoStop}%
\end{thebibliography}%

\end{document}